\documentclass[sigconf]{acmart}

\AtBeginDocument{%
  \providecommand\BibTeX{{%
    \normalfont B\kern-0.5em{\scshape i\kern-0.25em b}\kern-0.8em\TeX}}}

\copyrightyear{2021}
\acmYear{2021}
\setcopyright{rightsretained}
\acmConference[MICRO '21]{MICRO'21: 54th Annual IEEE/ACM International Symposium on Microarchitecture}{October 18--22, 2021}{Virtual Event, Greece}
\acmBooktitle{MICRO'21: 54th Annual IEEE/ACM International Symposium on Microarchitecture (MICRO '21), October 18--22, 2021, Virtual Event, Greece}
\acmDOI{10.1145/3466752.3480117}
\acmISBN{978-1-4503-8557-2/21/10}
\acmSubmissionID{977}
\acmBadgeL[https://www.acm.org/publications/policies/artifact-review-and-badging]{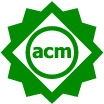}
\acmBadgeR[https://www.acm.org/publications/policies/artifact-review-badging]{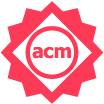}

\usepackage{fancyhdr} 
\usepackage{balance} 
\usepackage{xcolor,colortbl} 
\usepackage{graphicx} 
    \graphicspath{{./figures/}}
\usepackage{multirow} 

\makeatletter
\renewcommand{\sectionautorefname}{\S\@gobble}
\makeatother

  {\begin{list}{$\bullet$}%
     {\setlength{\parsep}{0pt}%
      \setlength{\topsep}{0pt}%
      \setlength{\leftmargin}{2pc}%
      \setlength{\itemsep}{2pt}}}%
  {\end{list}}

\newcommand{\hwfilter}{SquiggleFilter}

\newcommand{\ignore}[1]{}

\usepackage{float} 
\usepackage{listings} 
\begin{document}
\title{SquiggleFilter: An Accelerator for Portable Virus Detection} 

\author{Tim Dunn}
\email{timdunn@umich.edu}
\authornote{Both authors contributed equally to this research.}
\orcid{0000-0003-3429-4329}
\affiliation{%
 \institution{University of Michigan}
 \city{Ann Arbor}
 \state{MI}
 \postcode{48105}
 \country{USA}
}

\author{Harisankar Sadasivan}
\email{hariss@umich.edu}
\orcid{0000-0002-2832-458X}
\authornotemark[1]
\affiliation{%
 \institution{University of Michigan}
 \city{Ann Arbor}
 \state{MI}
 \postcode{48105}
 \country{USA}
}

\author{Jack Wadden}
\email{jackwadden@gmail.com}
\orcid{0000-0002-3055-3656}
\affiliation{%
 \institution{University of Michigan}
 \city{Ann Arbor}
 \state{MI}
 \postcode{48105}
 \country{USA}
}

\author{Kush Goliya}
\email{kgoliya@umich.edu}
\affiliation{%
 \institution{University of Michigan}
 \city{Ann Arbor}
 \state{MI}
 \postcode{48105}
 \country{USA}
}

\author{Kuan-Yu Chen}
\email{knyuchen@umich.edu}
\affiliation{%
 \institution{University of Michigan}
 \city{Ann Arbor}
 \state{MI}
 \postcode{48105}
 \country{USA}
}

\author{David Blaauw}
\email{blaauw@umich.edu}
\orcid{0000-0001-6744-7075}
\affiliation{%
 \institution{University of Michigan}
 \city{Ann Arbor}
 \state{MI}
 \postcode{48105}
 \country{USA}
}

\author{Reetuparna Das}
\email{reetudas@umich.edu}
\orcid{0000-0002-5894-8342}
\affiliation{%
 \institution{University of Michigan}
 \city{Ann Arbor}
 \state{MI}
 \postcode{48105}
 \country{USA}
}

\author{Satish Narayanasamy}
\orcid{0000-0001-5016-1214}
\email{nsatish@umich.edu}
\affiliation{%
 \institution{University of Michigan}
 \city{Ann Arbor}
 \state{MI}
 \postcode{48105}
 \country{USA}
}









\settopmatter{authorsperrow=4}


 \renewcommand{\shortauthors}{Dunn and Sadasivan, et al.}

\begin{abstract}

The MinION is a recent-to-market handheld nanopore sequencer. It can be used to determine the whole genome of a target virus in a biological sample. Its Read Until feature allows us to skip sequencing a majority of non-target reads (DNA/RNA fragments), which constitutes more than 99\% of all reads in a typical sample. However, it does not have any on-board computing, which significantly limits its portability.

We analyze the performance of a Read Until metagenomic pipeline for detecting target viruses and identifying strain-specific mutations. We find new sources of performance bottlenecks (basecaller in classification of a read) that are not addressed by past genomics accelerators. 

We present \hwfilter, a novel hardware accelerated dynamic time warping (DTW) based filter that directly analyzes MinION's raw squiggles and filters everything except target viral reads, thereby avoiding the expensive basecalling step. We show that our 14.3W 13.25$\text{mm}^{\text{2}}$ accelerator has 274$\times$ greater throughput and 3481$\times$ lower latency than existing GPU-based solutions while consuming half the power, enabling Read Until for the next generation of nanopore sequencers.

\end{abstract}
\maketitle

\section{Introduction} 
\label{intro}



The COVID-19 pandemic caused by the SARS-CoV-2 virus continues on a global scale. Today, diagnostic tests are widely available to detect SARS-CoV-2. Most of these tests involve some form of Polymerase Chain Reaction (PCR), a common technique for exponentially amplifying DNA/RNA. In order to detect a virus such as SARS-CoV-2, custom ``primers'' are first designed and manufactured which will only attach to and amplify specific regions of DNA/RNA in the target virus's genome. After PCR, the virus's presence or absence can be determined based on whether the amplification was successful or not. 


A significant shortcoming of the current approach is that PCR primers are targeted to a specific virus. {\bf Custom primer design is a complex, error-prone, and time-consuming process}~\cite{cdc_primer_messup}~\cite{covid_design_optimization}. Even though SARS-CoV-2's RNA was sequenced in early January 2020, validated SARS-CoV-2 specific PCR primers took several months to develop~\cite{covid_design_optimization}~\cite{ ARTIC_v3}. Lack of mass testing capability in the early stages of SARS-CoV-2 made it difficult to detect and control its spread, leading to a catastrophic pandemic. While we now have adequate testing capability for SARS-CoV-2, it is not unlikely for another novel virus like SARS-CoV-2 or its variants to emerge in the near future~\cite{future_path}, and if it does, we need to be prepared with adequate testing infrastructure in place to detect and control its spread in the early stages.

We envision a programmable virus detector (one that constructs whole viral genomes) that can be deployed worldwide. As soon as an emerging novel virus is discovered and sequenced, the reference genome of the novel virus would be distributed to all the devices, instantly turning them into targeted detectors. 

Our solution uses Oxford Nanopore Technologies' (ONT) MinION Mk1B (henceforth, referred to as the MinION), a new-to-market palm-sized DNA/RNA sequencer. It is fairly low-cost, portable, and can sequence long reads in real time.


\begin{figure}[h]
\centering
	\includegraphics[scale=0.02]{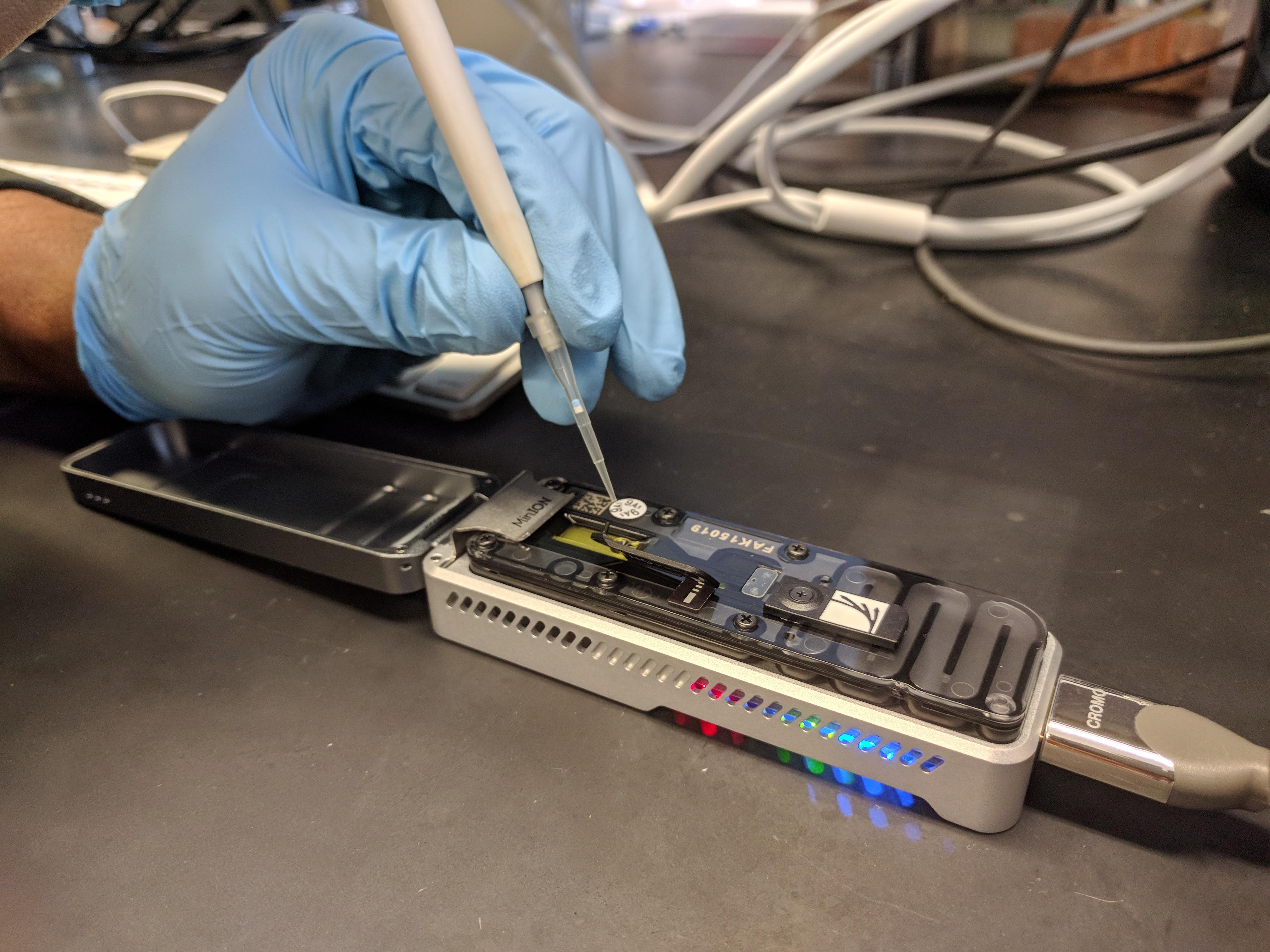}
	\vspace{-3mm}
  	\caption{MinION sequencer in our laboratory.}
  	\vspace{-2mm}
  	\label{fig:minion}
\end{figure}

We replace targeted PCR with universal PCR~\cite{random_primers}, which amplifies \emph{all} DNA/RNA. Thus, it avoids the problem of custom PCR primer design and deployment mentioned earlier. However, this introduces a different problem, as up to 99.99\% of the DNA/RNA in a typical biological specimen (e.g. saliva) is non-viral~\cite{minion_viral_seq} (non-target) and most belongs to the host. Amplifying all DNA/RNA preserves this ratio, resulting in the vast majority of sequencing and computing time and cost stemming from processing non-target DNA/RNA.

In order to solve this needle-in-a-haystack problem, ONT sequencers have a feature called \textbf{Read Until}~\cite{edwards2019real}. As reads (DNA/RNA fragments) are sequenced, they need to be analyzed in real-time. As soon as the computer classifies that the read is non-viral, the sequencer is instructed to eject it, which saves the time and cost of sequencing non-viral reads (>99\% of all reads). Unfiltered viral reads are used to construct the whole virus genome using reference-guided assembly (alignment and variant calling).



The MinION, however, does not have any on-board computing power to perform such secondary analysis. In this paper, {\bf we analyze the performance of the Read Until bioinformatics pipeline} for efficiently sequencing viral pathogens, and realize a portable computing solution that can be integrated with MinION. 

We discover new performance bottlenecks that are not addressed by past genomics accelerators~\cite{genax, genesis, darwin,seedex,gencache,nvm_read,genasm,wu2019fpga}. In particular, we find that the Deep Neural Network (DNN) basecaller (software that translates MinION's electrical squiggles to AGTC bases) dominates the computing time (~96\%). The aligner and variant caller, which have been the targets of recent accelerator research, constitute a much smaller fraction of compute. We also find that a current edge GPU is inadequate to keep up with the throughput of the MinION. Also, its high latency in classifying a read prevents us from taking advantage of the latency-critical Read Until feature of MinION.


\ignore{
Furthermore, sequencing portability and throughput is improving rapidly, as shown in Figure \ref{fig:portability}. ONT announced in 2019 that they are working with MinION prototypes with 16$\times$ greater sequencing throughput. Within the next few years, they hope to release a production flow cell with $100\times$ greater throughput~\cite{ont_tech_update}.
\begin{figure}[h]
\centering
	\includegraphics[scale=0.35]{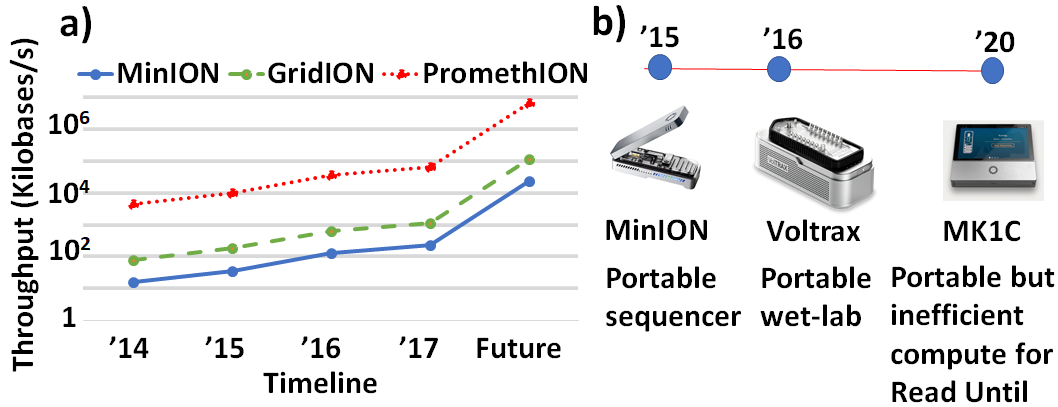}
  	\caption{a) Sequencing throughput is increasing exponentially b) Sequencing and wet-lab is portable. Compute, though portable, is insufficient for Read Until.}
  	\label{fig:portability}
\end{figure}
}

Converting squiggles to bases using a compute-intensive basecaller, and then aligning to check if a read belongs to the target virus is needlessly expensive for classifying it. Instead, \textbf{we skip the basecaller altogether by directly comparing each read's squiggles to the precomputed expected signal profile of the target virus's entire reference genome} (the ``reference squiggle''). By skipping the compute-intensive basecaller step, we improve efficiency significantly. 

{\bf We present \hwfilter, a hardware/software co-designed filter} which identifies non-target reads by directly comparing the real-time measured squiggles to the target virus's precomputed reference squiggle. A classification decision is made based on the degree of match. {\bf We develop a custom subsequence dynamic time warping (sDTW) algorithm}~\cite{dtw_alg} to perform this classification. It includes solutions that improve accuracy by adaptively examining longer read prefix lengths when needed. It also includes customizations that result in area efficient hardware.

sDTW-based \hwfilter~is significantly more efficient than a DNN-based basecaller, and its regular compute-bound characteristic makes it amenable for hardware acceleration. sDTW is a dynamic programming algorithm~\cite{dtw_alg_2} whose complexity is proportional to the product of the length of the reference (R) and query (Q). Its regular memory access pattern allows us to build a fast and space efficient 1D systolic array accelerator for sDTW with a constant number of processing elements. Fortunately, we find that almost all epidemic viruses have genome references of length 50,000 (R) bases or smaller (see Figure~\ref{fig:virus_genomes})~\cite{viral_genome_sizes}. As a result, our accelerator can easily complete the classification in $\sim$2R cycles (forward and backward of reference strand), and still meet the strict latency requirement for leveraging Read Until.

Our work makes the following contributions:
\vspace{-2mm}
\begin{itemize}\setlength\itemsep{0mm}
    \item we demonstrate that basecalling is the computational bottleneck in the virus sequencing pipeline. Read alignment and variant calling -- targets for prior accelerator work -- are not the bottleneck.
    \item we identify direct squiggle alignment (first proposed in~\cite{read_until_dtw}) as a more efficient alternative to basecalling and alignment when enriching low-concentration viral specimens with Read Until.
    \item we propose multi-stage sDTW and several modifications to vanilla sDTW to realize an accurate and efficient hardware accelerator.
    \item we co-design a sDTW hardware accelerator to filter non-viral reads, for variable read prefix and almost all infectious viral genome lengths
    \item we demonstrate that this hardware, unlike current approaches, will enable Read Until to scale with rapidly increasing nanopore sequencing throughput
    \item we quantify accuracy and efficiency of our classifier using real-world metagenomic datasets, including datasets collected from our wet-lab experiments for Read Until.
\end{itemize}

{\bf Results:} We design an edge device with compute capabilities similar to a Jetson Xavier System-on-Chip~\cite{xavier} consisting of \hwfilter, an edge GPU, and an 8-core ARM processor. We show that our proposed \hwfilter~can accurately distinguish target viral DNA/RNA from background human DNA/RNA. We evaluate accuracy using non-contagious lambda phage virus data sequenced in our own lab. 
In terms of efficiency, we show that our \hwfilter~accelerator has 274$\times$ higher throughput than the conventional software pipeline (using a basecaller) on an edge GPU while only consuming an area of 13.25$\text{mm}^2$ and power of 14.31W. \hwfilter's throughput is 233.65M samples/s, which far exceeds the maximum throughput of 2.05M samples/s on a MinION~\cite{minion_throughput}, and is adequate to handle up to a 114$\times$ increase in MinION's throughput in the future. The latency for classifying any read is 0.043ms, which is insignificant to Read Until decision's critical path. 

\section{Background}
\label{bg}

\subsection{Need for a Virus Detector}

\begin{figure}
    \centering
    \includegraphics[width=7cm]{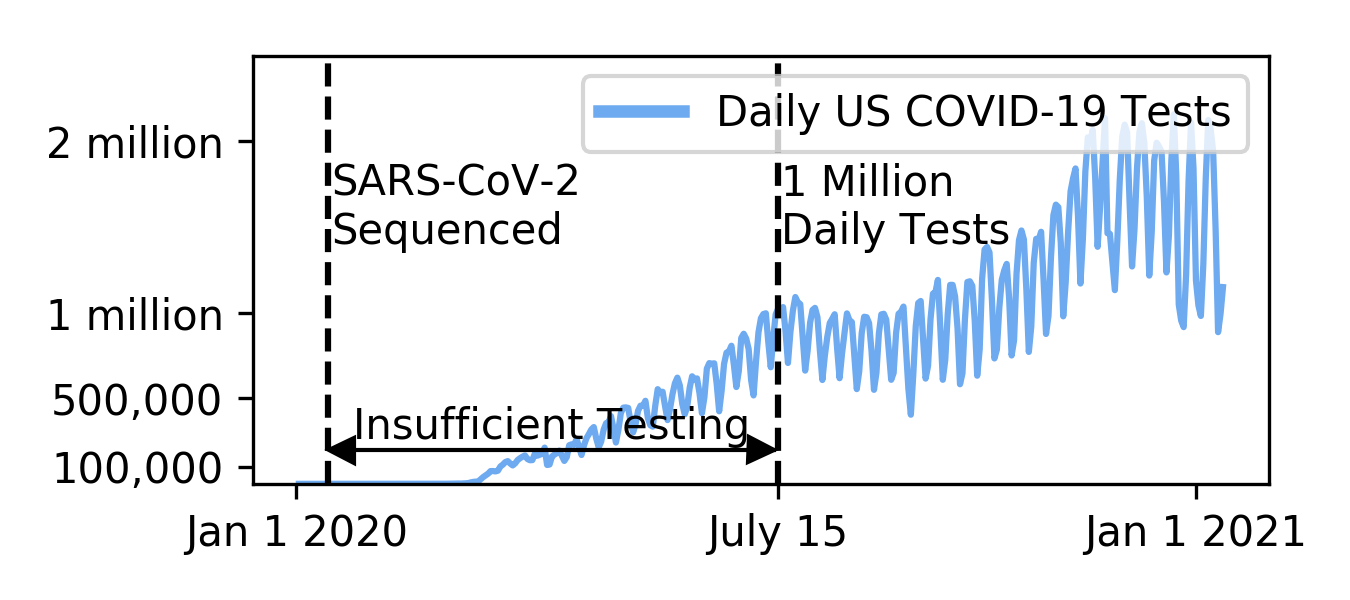}
    \vspace{-4mm}
    \caption{Progression of US COVID-19 testing~\cite{hasell2020cross}}
    \vspace{-4mm}
    \label{fig:covid19_testing}
\end{figure}

While SARS-CoV-2 was discovered -- and its RNA genome sequenced -- by early January 2020, it was not until several months later that mass testing was available worldwide. Figure~\ref{fig:covid19_testing} shows the steady increase in daily COVID-19 tests performed within the United States~\cite{hasell2020cross}. A widely established global testing infrastructure would have helped control the spread of the virus early on, possibly saving hundreds of thousands of lives. 

Given the increasing frequency of viral outbreaks, experts are concerned that it is only a matter of time before a new virus threatens the globe~\cite{future_path}. Thus, we need a virus testing technology that can be widely deployed \textit{ahead-of-time}, and reprogrammed to detect and identify mutations in novel viruses as soon as they emerge.






In this work, we focus on controlling the spread of novel infectious viruses in their early stages, as soon as they are discovered and sequenced. {\bf Our goal is to enable a universal rapid test that can determine the whole genome of a target virus using reference-guided assembly}. Targeting a specific virus enables us to make significant optimizations that help us reduce time and cost of sequencing and compute.

\subsection{State-of-the-art Virus Detectors}
\label{sec:prior-tests}
\label{priordetectors}
\begin{table}[htb]
\renewcommand{\arraystretch}{1.2}
  \centering
  \footnotesize
  \begin{tabular}{ccccc}
    \toprule
    \shortstack{\textbf{Tests}\\\ }  & \textbf{\shortstack{Diagnostic\\Power}}  & \shortstack{\textbf{Programmable}\\\ } & \shortstack{\textbf{Time}\\(min)} & \shortstack{\textbf{Cost}\\(\$)}\\
    \midrule
    \multicolumn{5}{l}{\textbf{Antigen-based test}}   \\
    Paper \cite{abott_paper} & presence  &  & 15 & 5 \\
    \midrule
     \multicolumn{5}{l}{\textbf{Non-sequencing molecular test}}   \\
     RT-LAMP \cite{lamp_cite}\cite{neb_lamp} & presence & & 60 & 15 \\
    RT-PCR \cite{rtpcr_primers} & presence & & 120-240 & $<$10\\
     \midrule
     \multicolumn{5}{l}{\textbf{Sequencing based molecular test ($30\times$ coverage)}}   \\
    ARTIC \cite{pcr_cdna}\cite{rtpcr_primers} & 98 targets & & 305 & 100  \\
    LamPORE \cite{lampore} & 3 targets &  & $<$65 & -NA-\\
    \textbf{RNA}: 1\% virus & whole genome & \checkmark & 240 & 110\\
    \hspace{10mm}0.1\% virus \cite{drna_kit} & whole genome & \checkmark & 1206 & 190 \\
    \textbf{DNA}: 1\% virus & whole genome & \checkmark & 320 & 105 \\
    \hspace{10mm}0.1\% virus \cite{cdna_kit} & whole genome & \checkmark & 470 & 120\\
     \bottomrule
   \end{tabular}
   \vspace{1mm}
  \caption{A comparison of popular commercial and ONT sequencing-based virus detectors for SARS-CoV-2.}
  \label{tab:most_popular_covid_tests}\vspace{-2mm}
\end{table}

Table \ref{tab:most_popular_covid_tests} lists commonly used tests and ONT-based sequencing solutions for SARS-CoV-2. None of the methods except direct RNA or DNA sequencing are programmable, and therefore, are not effective in controlling the pandemic in its early stages. Antigen (paper) tests detect specific surface proteins on the virus. They are cheap, portable, and fast. However, they have low sensitivity and can only detect viruses present at high concentrations.

Molecular tests identify specific regions of interest in a virus's genome and amplify this DNA if present in the specimen. Polymerase Chain Reaction (PCR) is a common technique used for amplification. It has high sensitivity~\cite{covid_test_survey} but requires thermal cycling, which can be slow and expensive. LAMP (Loop Mediated Isothermal Amplification) is a more recent technology that obviates the need for a thermal cycler, but its primers are more complicated to design than PCR. 

If amplification was successful (i.e., target DNA is present), it can be detected using fluorometry or colorimetry. Most clinical tests for SARS-CoV-2 stop here. However, by sequencing the amplified specimen, we can assemble portions of virus's genome, depending on the number of targets amplified. ARTIC~\nocite{ARTIC_v3_protocol} and LamPORE~\cite{lampore} amplify 98 and 3 genes respectively, and then use ONT's nanopore sequencing. 

Current solutions for virus detection use multiplex primer sets specific to a virus. Primer design is a complex, error-prone and time-consuming process~\cite{cdc_primer_messup}~\cite{covid_design_optimization}. Thus, they are not an effective solution for early pandemic control. The COVID-19 pandemic highlights this problem, where designing and distributing target-specific primers was challenging, especially when supply chains broke amidst the pandemic. 

An alternative to developing custom primers is to directly sequence the specimen following amplification with universal primers, which non-selectively amplify all DNA. This amplification step is required to increase the quantity of DNA, which greatly reduces average capture time (the time required for a DNA strand to enter a nanopore) and therefore sequencing time. 
The wet-lab protocol followed, Sequence Independent Single Primer Amplification (SISPA)~\cite{sispa_protocol,sispa_ont}, is universal and hence can be used on all RNA viruses. SISPA has four major steps: (1) RNA extraction, (2) complementary DNA generation, (3) PCR amplification, and (4) final sequencing specimen preparation. 

\ignore{
\begin{figure*}[h]
    \centering
	\includegraphics[width=\textwidth, scale=0.10]{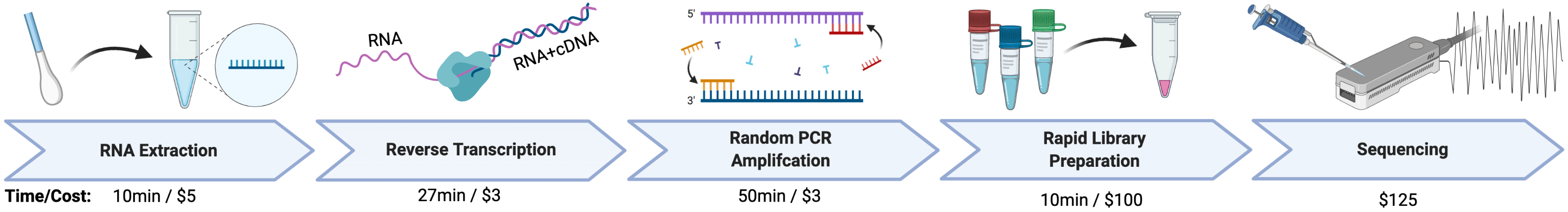}
	\vspace{-6mm}
  	\caption{Recommended end-to-end RNA preparation, amplification, and sequencing pipeline. All times and costs are modeled based on in-house wet-lab experience, and reagents and associated protocols. Sequencing time depends on host-to-bug ratio, \%healthy pores, capture time, and Read Until.}
  	\label{fig:wet-lab}
\end{figure*}
}

A significant hurdle to SISPA-based sequencing is that following amplification, the specimen contains the genetic material of the target virus among a sea of human and bacterial DNA/RNA. The proportion of target virus DNA/RNA can be as low as 0.01\% percent~\cite{minion_viral_seq}. As a result, the time and cost of sequencing and data processing for this approach is significantly greater than that of custom primer-based solutions. 

If this cost barrier can be overcome, this approach would enable detection of novel viruses without requiring months to develop and distribute virus-specific primers. Read Until can greatly increase the efficiency of sequencing by filtering out non-target reads using the virus's reference genome. Current Read Until approaches are limited by insufficient throughput, but our hardware accelerated \hwfilter~ensures the future scalability of Read Until on higher throughput sequencers.


\subsection{Portable MinION Sequencer}	
\label{minION}


Oxford Nanopore Technology's (ONT) MinION offers multiple benefits that makes it a uniquely attractive solution for mobile and rapid virus detection. 

\textbf{Long reads:} MinION sequencers are capable of measuring long strands of DNA, and can theoretically sequence any strand, regardless of length. The current world record stands at over 4 million bases~\cite{read_len_world_record}.

\textbf{Cost:} The MinION only costs \$1,000, and offers affordable specimen preparation kits (\$100/use) and flow cells (\$125/use assuming 4$\times$ re-use). In comparison, it costs \$80,000-\$100,000 to purchase even the most affordable ``Next Generation Sequencing'' machines.

\textbf{Real-time:} MinION sequencers provide real-time, streaming output from the device. Streaming signal output enables on-the-fly secondary analyses, and the ability to stop sequencing as soon as the desired coverage is reached.


\textbf{Portability:} A key feature that sets the MinION sequencer apart from all other sequencers in terms of wet-lab, sequencing and compute as shown in Figure~\ref{fig:portability}. The portable compute, however, remains inefficient for real-time sequencing.
\begin{figure}[th]
    \centering
	\includegraphics[width=\columnwidth]{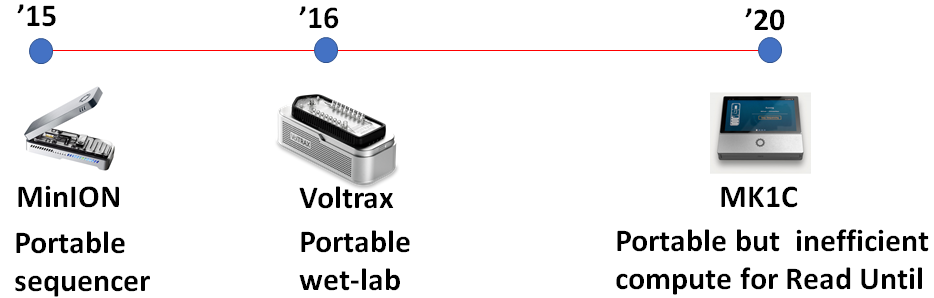}
	\vspace{-0.5cm}
  	\caption{Sequencing and wet-lab is portable. Compute, though portable, is insufficient for Read Until.}
  	\label{fig:portability}
\end{figure}

\sloppypar{\textbf{Target enrichment:} 
An especially exciting capability of the MinION sequencer is {\bf ``Read Until"}, which ejects non-target DNA/RNA strands by reversing the electrical potential across the pore. This effectively enables digital enrichment of target DNA/RNA in low-concentration specimens.}


However, a slow read classification results in wasted sequencing time. Currently, the MinION has no inbuilt computing power to make Read Until decisions. We additionally find that commodity GPUs are undesirable in terms of both throughput, latency and power.

\begin{figure*}[th]
    \centering
	\includegraphics[scale=0.40]{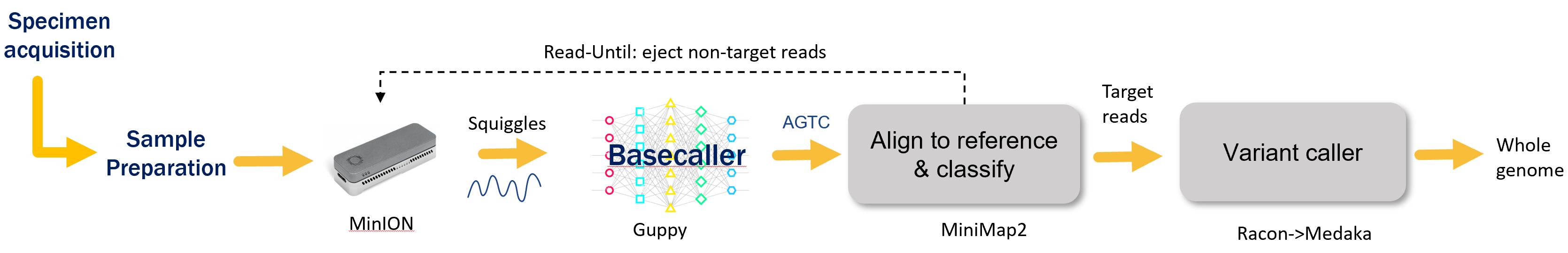}
	\vspace{-4mm}
  	\caption{A Read Until pipeline for targeted reference-guided assembly of a virus genome.}
  	\label{fig:basesw}
  	\vspace{-2mm}
\end{figure*}

\begin{figure}[h]
    \centering
	\includegraphics[width=7.5cm]{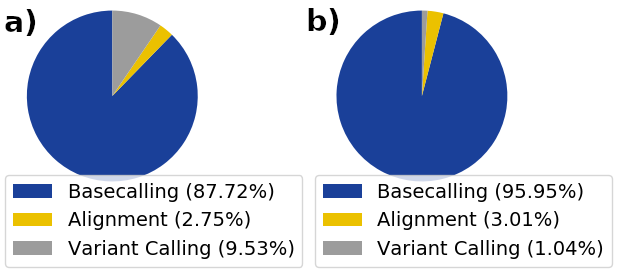}
	\vspace{-3mm}
  	\caption{Basecalling is the bottleneck in a Read Until assembly of a SARS-CoV2 genome from specimens with \textbf{a)} 1\%, and \textbf{b)} 0.1\% viral reads.}
  	\label{fig:pipeline}
  	\vspace{-2mm}
\end{figure}

\section{Compute Bottlenecks in Portable Virus Detection}	
\label{basesw}

Our goal is to build a cost and time efficient sequencing pipeline for determining the whole genome of a targeted virus, but without using custom primers for target amplification. We seek to reduce time and cost using the Read Until feature of Oxford Nanopore (ONT)'s palm-sized MinION sequencer. 

To this end, we constructed a software pipeline using state-of-the-art bioinformatics tools and analyzed its performance. Our profiling results expose new performance bottlenecks that are different from those targeted in  past accelerators for human genome sequencing~\cite{genax, genesis, darwin,seedex,gencache,nvm_read,genasm,wu2019fpga}.

\subsection{Bioinformatics Pipeline} 

The MinION sequencer measures electrical current signals that represent the bases (A, G, T, C) moving through each pore, recording approximately 10 samples for each base. All the active pores (up to 512 in the MinION) concurrently produce squiggles for the reads flowing through them. These squiggles can be analyzed in real-time as the reads flow through the pores.

Figure~\ref{fig:basesw} illustrates the analysis pipeline for the squiggles. A {\em basecaller} translates squiggles into bases. The latest basecallers (such as ONT's Guppy~\cite{wick2019}) use compute-intensive DNNs, which must be large and deep to attain state-of-the-art accuracy. Guppy processes reads in chunks of 2000 samples, and uses five bidirectional LSTM layers for encoding followed by a custom CTC (Connectionist Temporal Classification) decoder. ONT provides two versions of its basecaller: a high-accuracy version (Guppy), and another that trades off accuracy for performance (Guppy-lite). 

In our Read Until pipeline, squiggles of a read are basecalled in real-time. After a short prefix of a read has been basecalled, it is then processed by an aligner (MiniMap2~\cite{li2018minimap2}) that aligns the read to the target's reference genome. If a good alignment is found, then the read is classified as a target and passed on to the next stage. Otherwise, a signal is sent to the MinION device, instructing it to eject the non-target read from further sequencing. {\bf Thus, the critical computing path for Read Until includes both the basecaller and aligner}. 

The target reads are collected and analyzed by a variant caller (Racon~\cite{racon} followed by Medaka~\cite{medaka}). We seek to cover every position in the reference genome by 30 reads (30$\times$ coverage). The variant caller analyzes the reads piled up at each reference genome location, and identifies any genomic differences (``variants'') between the sequenced and reference viruses. 
{\bf As the variant caller is not involved in Read Until decisions, it is off the critical path}.

\subsection{Performance Bottlenecks}
\label{sec:need}

Figure~\ref{fig:pipeline} shows the performance bottlenecks of the bioinformatics pipeline (Section~\ref{basesw}) used to assemble the whole SARS-CoV2 genome, evaluated on the CPU and GPU in Table~\ref{tab:devices}. The results are shown for two representative biological specimens, one where the target viral reads constitute 1\% of all the reads, and the other 0.1\%. 

{\bf We observe that a large fraction of computing time (96\%) goes towards basecalling}. This is in spite of using the more efficient, but less accurate, Guppy-lite.  


Compute spent towards aligning (MiniMap2) and variant calling (Racon and Medaka) constitutes significantly smaller fraction, especially for specimens with low viral load (0.1\%). In contrast, prior work on genomics accelerators targeted aligners and variant callers used for reference-guided assembly of human DNA~\cite{genax, genesis, darwin,seedex,gencache,nvm_read,genasm,wu2019fpga}. There are several reasons for this significant difference, discussed next.

All the reads are aligned to a target viral genome to classify them as target or non-target. This alignment step, however, is significantly less compute intensive compared to aligning to a human genome, because  viral genomes are much shorter ($\approx$30,000 bases) than human DNA (3 billion bases).

Only a small fraction of target reads (1\% to 0.1\%) need to be processed for reference-guided assembly of a viral genome. Therefore, the variant caller is invoked only for a small fraction of sequenced reads. Also, given that viral genomes are shorter, we find that the variant caller does not consume much compute resources. Furthermore, the variant caller is not on the critical path for using Read Until, as it is not required for classifying reads.




We find that even a 250W Titan GPU has barely enough basecalling throughput (with low accuracy Guppy-lite) to keep up with a MinION's maximum sequencing throughput. An edge GPU (e.g., Jetson Xavier's) is several times slower than that, and therefore it cannot process all the sequenced reads in real-time to exploit the latency sensitive Read Until feature.

\begin{figure}[h]
    \centering
	\includegraphics[width=\columnwidth]{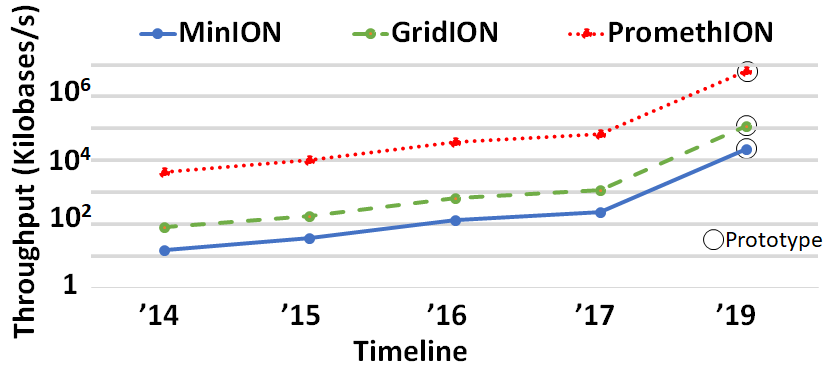}
	\vspace{-0.7cm}
  	\caption{Sequencing throughput is increasing exponentially~\cite{trans_rates}.}
  	\label{fig:throughput_trend}
\end{figure}

Sequencing throughput, however, continues to grow, as shown in Figure~\ref{fig:throughput_trend}. Oxford Nanopore Technologies (ONT)'s GridION is only slightly larger, but has 5$\times$ the sequencing throughput of a MinION. ONT announced in 2019 that they are working with MinION prototypes that provide 16$\times$ sequencing throughput of MinION devices available in the market today. Within the next few years, they plan to release a production flowcell with $100\times$ greater throughput~\cite{ont_tech_update}. 

Currently, the MinION does not have any on-board compute capability. Our goal is to map all the secondary compute analysis onto an edge system-on-chip so that it can be integrated with the MinION. We address this growing computing need with our small, low-power accelerated \hwfilter, which greatly reduces the basecalling and alignment computation required for non-target reads.



\section{SquiggleFilter: A Squiggle-level Targeted Filter using Dynamic Time Warping}
\label{sw}

As discussed in Section~\ref{basesw}, classifying a read being sequenced by analyzing its short prefix as target or not, in real-time, is the compute bottleneck. Additionally, basecalling for this classification consumes the most compute time.

Instead of using a basecaller (DNNs) and MiniMap2 aligner to classify a read's prefix, we discuss \hwfilter's~algorithm that directly aligns each read's electrical signals (query) to the target viral genome's precomputed electrical signal (reference). As a majority of the reads are non-targets, we reduce latency and save much of the work done in basecalling and aligning these non-target reads. 

\hwfilter~aligns the query squiggle with a precomputed reference squiggle of the viral genome using a variant of the dynamic time warping (DTW) algorithm~\cite{han}. Recent work has eschewed sDTW due to it's $\Theta(NM)$ complexity~\cite{ru_lec, rubric_new, rubric, kovaka}, but we demonstrate that since both queries (read prefixes) and virus genomes are short, it is a practical solution for viral read enrichment. We further demonstrate its effectiveness on real sequencing data for a SARS-CoV-2 specimen. 

Finally, we propose multi-stage sDTW filtering to improve efficiency, and discuss several improvements to conventional sDTW that help realize an efficient hardware accelerator.

\subsection{Constructing the Reference Squiggle}
\label{sec:ref_squiggle}

\begin{figure}[h]
    \centering
	\includegraphics[width=7cm]{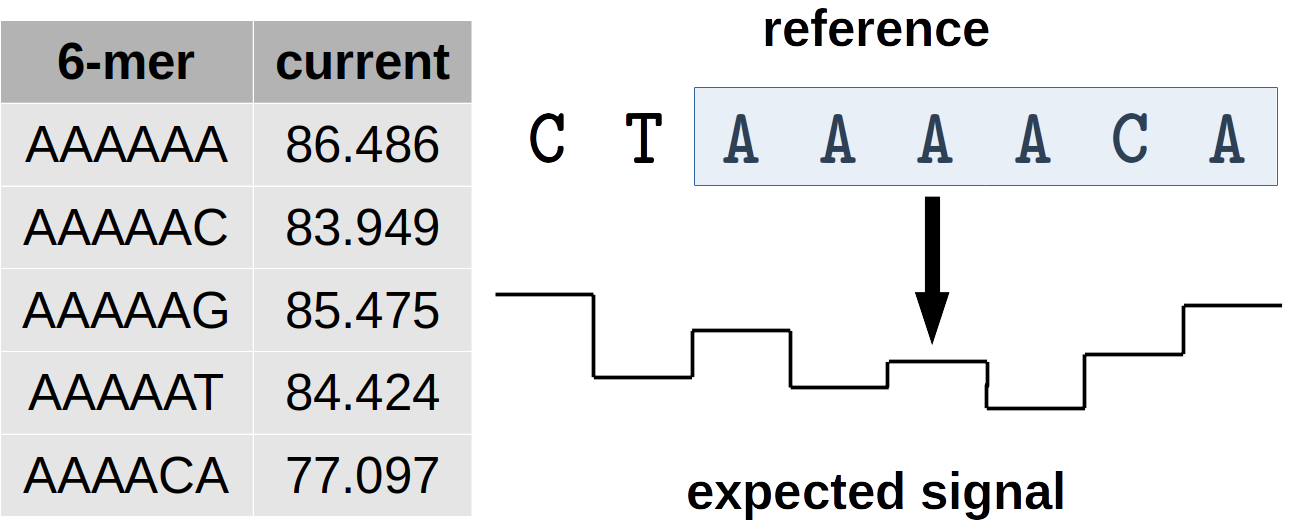}
	\vspace{-3mm}
  	\caption{Aligning reference bases to expected currents.}
  	\label{fig:kmer_model}
\end{figure}

In order to align raw signals to a reference genome, the known sequence of bases must first be converted to an expected current profile~\cite{tombo,read_until_dtw,nanopolish}. As a strand of DNA passes through a nanopore, the current measured is affected by 5-6 adjacent bases simultaneously. A lookup table is provided by ONT which contains the expected current (in pA) for every possible combination of six bases (``6-mer'')~\cite{kmer_models}. This conversion is demonstrated in Figure \ref{fig:kmer_model}, after which the expected signal is normalized using the mean and standard deviation.

\subsection{Normalizing Query Squiggles}
\label{sec:norm}

\begin{figure}[h]
    \centering
	\includegraphics[width=8cm]{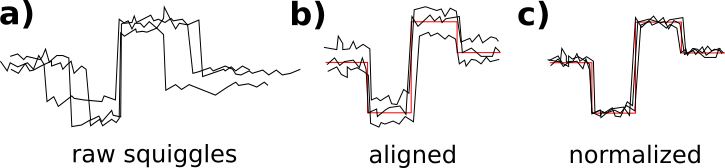}
	\vspace{-2mm}
  	\caption{\textbf{a)} Three raw current measurements (``squiggles'') for the same sequence of bases. We then show squiggles aligned to the expected signal \textbf{b)} without, and \textbf{c)} with normalization.}
  	\label{fig:squiggle_norm}
\end{figure}

Figure~\ref{fig:squiggle_norm}a shows a contrived minimal example of multiple raw nanopore signals corresponding to the same sequence of bases.
Due to a variable rate of DNA/RNA translocation through the nanopore, these signals are out-of-sync (transitions between current levels do not occur simultaneously).
Using Dynamic Time Warping (discussed next) solves this issue, and signals are aligned to the expected signal profile (shown in red in Figure~\ref{fig:squiggle_norm}b).
Slight differences in applied bias voltages at each nanopore cause the measured currents to differ significantly, which is why normalization within each read is additionally helpful (Figure~\ref{fig:squiggle_norm}c).

\subsection{Subsequence Dynamic Time Warping}
\label{sec:sdtw}

Dynamic Time Warping (DTW) is a dynamic programming algorithm which is commonly used to align out-of-sync signals~\cite{dtw_alg,keogh2003}.
Our filter applies subsequence DTW (sDTW), a slight modification of standard DTW which allows the entire query signal to align to any small portion of the reference, rather than forcing end-to-end alignment of both sequences.

The original sDTW algorithm works as follows for subsequence query $Q$ of length $N$, reference sequence $R$ of length $M$, and scoring matrix $S$:
\begin{lstlisting}
def sDTW(Q,R):
    S = zeros(N,M)
    S[0,0] = (Q[0]-R[0])$^\mathrm{\texttt{2}}$
    for i in range(1,N):
        S[i,0] = S[i-1,0] + (Q[i]-R[0])$^\mathrm{\texttt{2}}$
    for i in range(1,N):
        for j in range(1,M):
            S[i,j] = (Q[i]-R[j])$^\mathrm{\texttt{2}}$ + min(
                S[i-1,j-1], S[i,j-1], S[i-1,j])
    return min(S[N,:])
\end{lstlisting}

\begin{figure}[h]
    \centering
	\includegraphics[width=8.5cm]{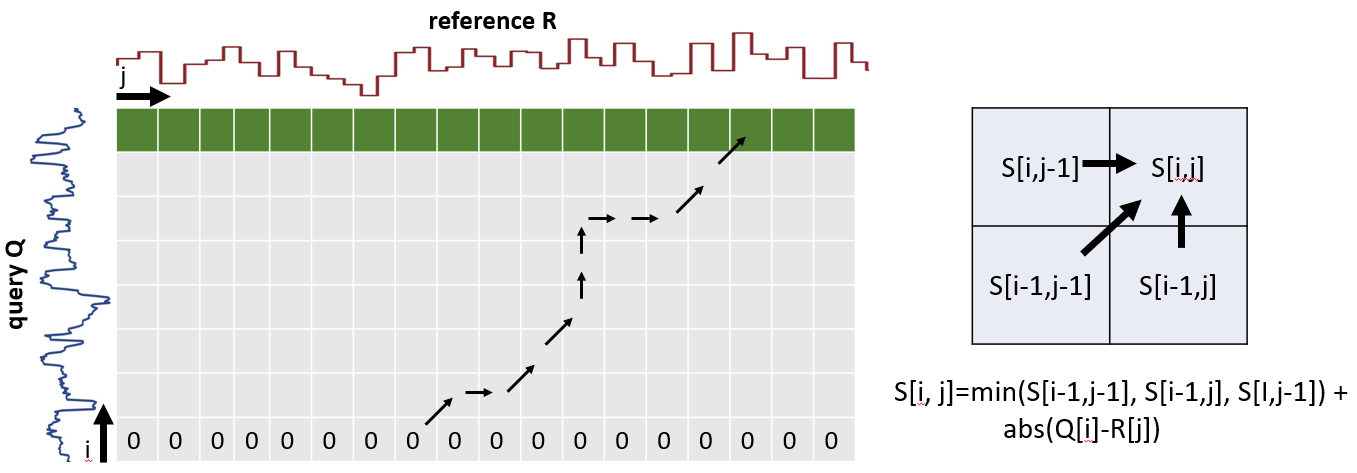}
	\vspace{-7mm}
  	\caption{Dynamic time warping algorithm.}
  	\label{fig:sdtw}
\end{figure}
The above algorithm dynamically computes all possible alignments of the query $Q$ to reference $R$ (keeping only the best ones) while allowing arbitrary many-to-one or one-to-many mappings between the two signal profiles. It is illustrated in Figure~\ref{fig:sdtw}.
Matrix $S$ records a running tally of the net squared differences between the two signals (using the best alignment of $Q[0:i]$).
At the end, $S[N,j]$ (highlighted top row in Figure~\ref{fig:sdtw}) contains the alignment cost of $Q$ to a subsequence of the reference $R[x:j]$, where $x$ is the start of the best alignment ending at $j$.
The minimum value in this row corresponds to the least squared difference in signal between alignments of the signal to the reference, and thus the cost of the optimal alignment.

\subsection{sDTW for Virus Detection}
\label{sec:virus}

\begin{figure}[h]
    \centering
	\includegraphics[width=8.5cm]{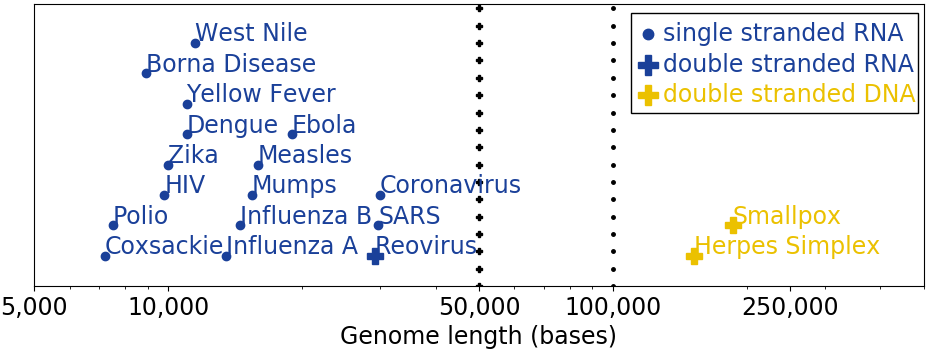}
	\vspace{-0.8cm}
  	\caption{Epidemic virus genome lengths.}
  	\label{fig:virus_genomes}
\end{figure}

The majority of viruses which are responsible for human epidemics have relatively small single-stranded RNA genomes~\cite{viral_genome_sizes}, as is demonstrated in Figure~\ref{fig:virus_genomes}.
The two notable exceptions are \textit{smallpox} and \textit{herpes simplex}, which have larger and more chemically stable double-stranded DNA genomes. Because most viruses have small genomes, we design our filter to operate on viruses with single-stranded genomes of length less than 100,000 bases. Equivalently, the filter works on viruses with double-stranded genomes less than 50,000 bases long. At such short reference genome lengths, it is computationally feasible to compare reads to the entire reference genome for filtering. This would not be a feasible solution for complex organisms such as humans, with genomes approximately 3 billion base pairs long.

\subsection{sDTW is an Effective Filter}
We seek to design a solution that is capable of detecting all strains of a particular viral species. It is therefore important that our filter is tolerant to variants in the sequenced genome relative to the reference genome used by our filter. We found that reference-guided filtering can be accurate regardless of viral strain, since the number of mutations between different strains is low. Table \ref{tab:covid_strains} presents the number of single base mutations between an assembled virus genome for several known SARS-CoV-2 strains, relative to the original Wuhan reference assembly~\cite{zhou2020pneumonia}. No insertions or deletions were observed. Strains were defined using NextStrain's~\cite{nextstrain} classification of all sequenced SARS-CoV-2 genomes into groups of shared ancestors, or ``clades'', and data was sourced from the GISAID database~\cite{gisaid}.

\begin{table}[htb]
  \centering
  \footnotesize
  \renewcommand*{\arraystretch}{1.2}
  \begin{tabular}{ccccc}
    \toprule
    \textbf{Clade} & \textbf{Mut.} & \textbf{GISAID ID} & \textbf{Lab of Origin} & \textbf{Country} \\
    \midrule
    19A & 23 & 593737 & SE Area Lab Services & Australia \\
    19B & 18 & 614393 & Bouake CHU Lab & Ivory Coast \\
    20A & 22 & 644615 & Dept. Clinical Microbiology & Belgium \\
    20B & 17 & 602902 & NHLS-IALCH & South Africa \\
    20C & 17 & 582807 & Public Health Agency & Sweden \\
    \bottomrule
   \end{tabular}
  \caption{There are few mutations between SARS-CoV-2 strains, relative to the Wuhan reference genome.}
  \label{tab:covid_strains}\vspace{-2mm}
\end{table}

\begin{figure}[h]
    \centering
	\includegraphics[width=8cm]{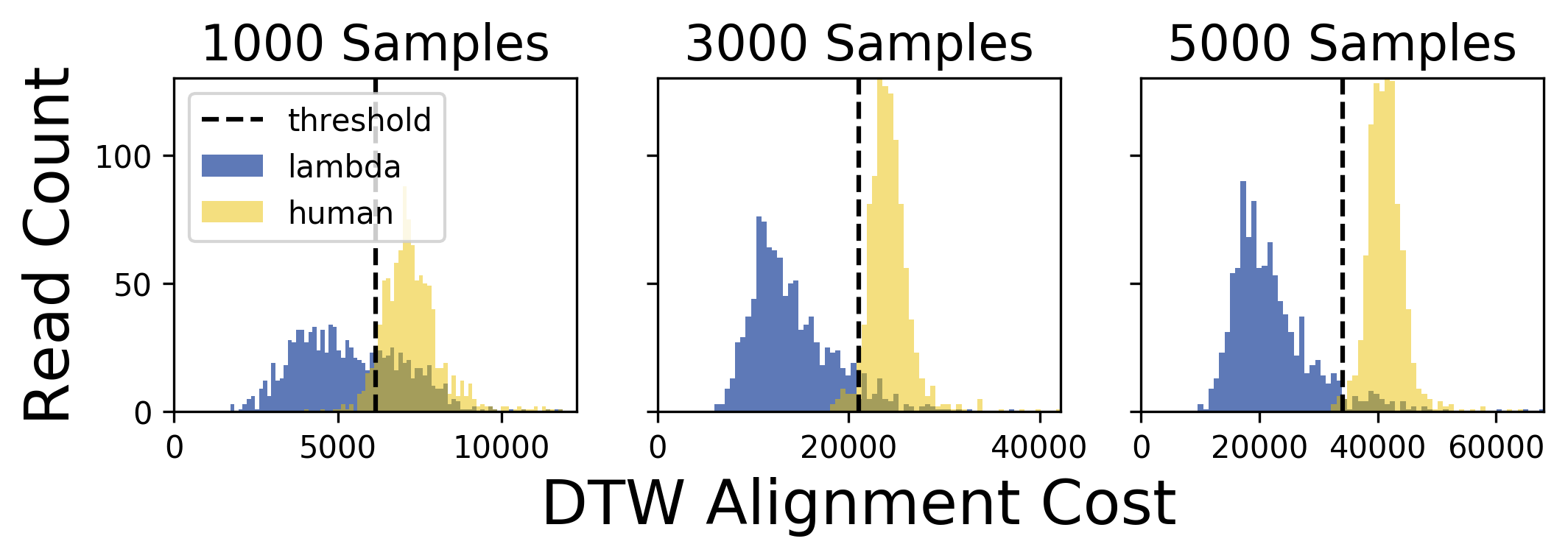}
	\vspace{-3mm}
  	\caption{sDTW cost distributions for reads of 3 prefix lengths, aligned to the lambda phage genome.}
  	\label{fig:dtw_hist}
\end{figure}

Since there are only a handful of mutations between various SARS-CoV-2 strains, the final sDTW alignment cost will not be significantly impacted. This cost is used to determine whether a given read aligns to the viral reference genome by comparing it to a constant threshold. If the alignment cost exceeds the chosen threshold, then the squiggle did not match well with any subsequence of the reference genome's expected current profile, and the read can be discarded.
Figure~\ref{fig:dtw_hist} shows that a static threshold can be used to distinguish between viral and human DNA fragments (discarding reads above the threshold and keeping reads below the threshold) even when only a few thousand signals have been captured. Due to the slight overlap in final alignment costs, some reads will be incorrectly classified when using a static threshold.

\subsection{Multi-stage sDTW Filtering}
\label{multi}

We observed that as a read's sequenced prefix length increases, the sDTW alignment cost is more accurately able to distinguish between target and non-target DNA (there is a decrease in overlap between cost distributions in Figure~\ref{fig:dtw_hist}). However, waiting to make a Read Until decision increases the proportion of non-target DNA sequenced.

Therefore, instead of a single-stage filter that chooses a constant read length and threshold, we can filter in multiple stages. The first stage examines a shorter read length (e.g. 1000 samples), but chooses a less aggressive threshold that may let many non-target reads through. Non-target reads filtered and ejected using Read Until at this stage would be very short. If a read is retained, it is sequenced further. The second stage then examines the longer read prefix (e.g. 5000 samples), and filters using a more aggressive threshold. Intermediate results can be stored to avoid recomputation. In this way, several stages enable the classifier to filter a majority of non-target reads after seeing only a short prefix. Only reads with initial low-confidence are sequenced more before a decision is made. We have designed our hardware accelerator with this (optional) capability.

\subsection{sDTW Algorithm Improvements}
\label{sec:alg}
We propose several modifications to sDTW which help improve either our accelerator's efficiency or accuracy of non-target read filtering.

\textbf{Absolute Difference:} We reduce hardware area and avoid multiplication by using \texttt{abs(Q[i]-R[i])} as our distance metric instead of \texttt{(Q[i]-R[j])$^\mathrm{\texttt{2}}$}.

\textbf{Integer Normalization:} Our solution uses 8-bit fixed point arithmetic during normalization, with no significant impact to classification accuracy (see Figure \ref{fig:alg_mods}).

\sloppypar{\textbf{No Reference Deletions:} Since the MinION averages 10 samples per base pair, it is unnecessary during sDTW computation for a single squiggle value to be able to align to multiple bases. We removed the possibility of reference deletions entirely from our dynamic programming computation, so that \texttt{S[i,j] = abs(Q[i]-R[j]) + min(S[i-1,j-1], S[i-1,j])}.}

\textbf{Match Bonus:} This final modification improves filtering accuracy. We found that reads with higher average translocation rates generally have higher alignment costs. To ensure sDTW alignment costs solely represent quality of alignment and are independent of translocation rate, we implemented a ``match bonus'' that rewards reads for matching additional reference bases, reducing the alignment cost for each matching base by a constant (10) scaled by the number of signals aligned to the previous reference base (thresholded to 10).

\subsection{Need for an Accelerator}
Despite the reduction in computation when compared to basecalling, sDTW alignment is still too slow to run on commodity hardware. sDTW alignment does avoid expensive floating point operations, instead requiring 8-bit integer comparisons and additions/subtractions. sDTW also has a smaller memory footprint (60,000 reference bases) compared to Guppy-lite (284,000 weights) when filtering SARS-CoV-2. Despite memory and operation complexity advantages, however, the number of operations required for sDTW (1,400 million) is greater than that of Guppy-lite (141 million). This is still more efficient than Guppy (2,412 million). In order to meet current and future MinION device requirements for Read Until, it is necessary to design an accelerator.
\section{Accelerated SquiggleFilter}
\label{hw}

\begin{figure}[h]
    \centering
	\includegraphics[width=\columnwidth]{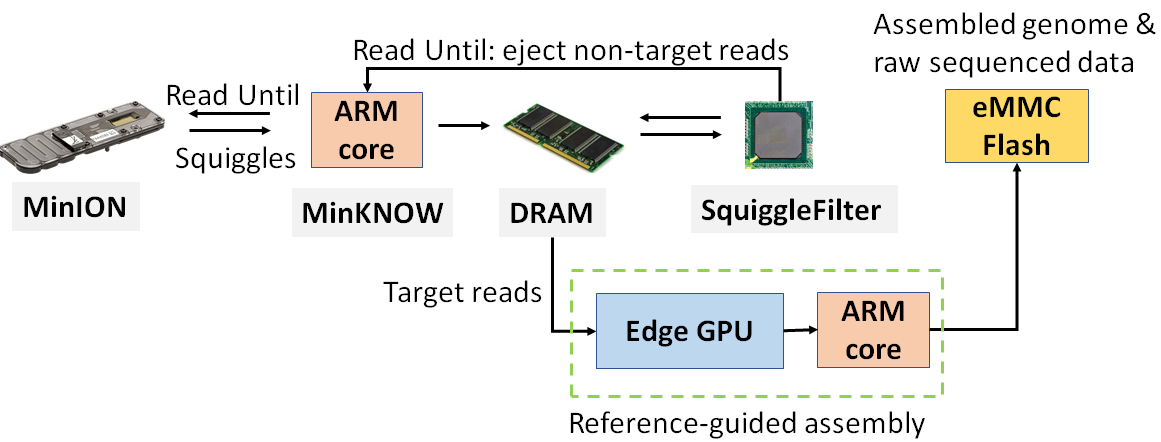}
	\vspace{-0.8cm}
  	\caption{System-on-Chip design with the accelerated hardware filter on ASIC integrated with NVIDIA GPU and 8-core ARM v8.2 64-bit CPU}
  	\label{fig:arch_top}
\end{figure}

We present a System-on-Chip for reference-guided assembly of target viruses, shown in Figure~\ref{fig:arch_top}. Its capabilities are similar to a Nvidia Jetson TX2, except for our \hwfilter~accelerator. Our \hwfilter~accelerator classifies and filters non-target reads, which constitute >99\% of all reads in most biological specimens. Thus, a large fraction of computing identified in Section~\ref{basesw} is handled by our \hwfilter~accelerator. Furthermore, our accelerator enables low latency read classification, allowing us to use Read Until to eject non-target reads after sequencing only a short prefix.

Target reads (and any false positives) are processed off of Read Until's critical path. Only these small fraction of reads need to be basecalled, aligned, and variant called. We find that we can perform these computations on an edge GPU (basecaller) and ARM processor (aligner and variant caller), and still construct the whole viral genome in approximately 10 minutes. Unfiltered non-target reads (false positives due to sDTW algorithm) will fail to align to the viral reference genome after basecalling, and so they will be discarded without affecting the accuracy of conventional reference-guided assembly. The final assembled genome and raw sequencing data is written to a 32GB eMMC 5.1 flash memory, which is sufficient to store one day's worth of sequencing data. 

We now present the 1D systolic array based \hwfilter~accelerator for our squiggle-level classification algorithm discussed in Section~\ref{sw}. It can be programmed to target any novel viral genomes less than 100K bases. It supports variable query length. That is, it can classify read prefixes of different lengths, and thereby supports multi-stage filtering. The size of the systolic arrays and buffers are derived from our analysis of real-world metagenomic data.

\ignore{\begin{figure}[h]
    \centering
	\includegraphics[scale=0.45]{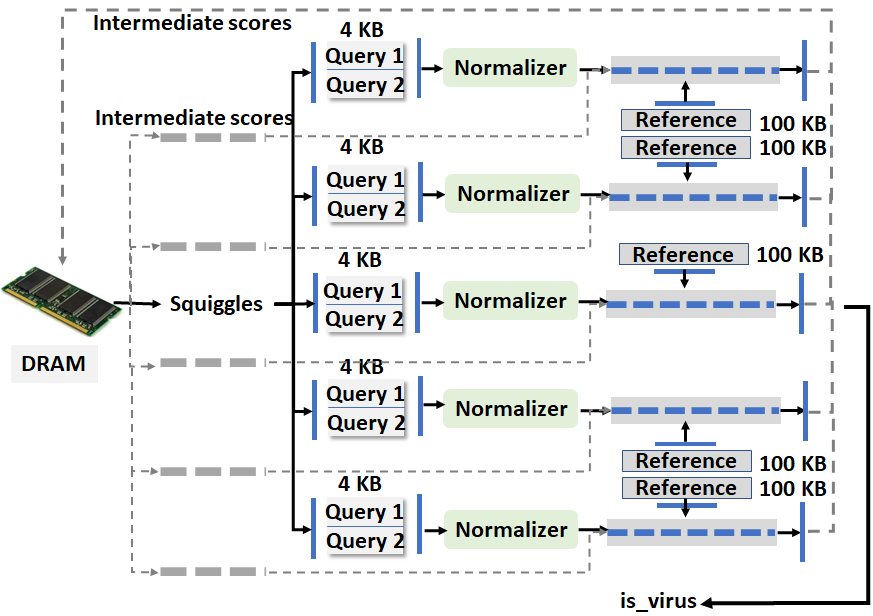}
  	\caption{\hwfilter~Accelerator. }
  	\label{fig:tiles}
\end{figure}}

\begin{figure*}[h]
    \centering
	\includegraphics[width=17cm]{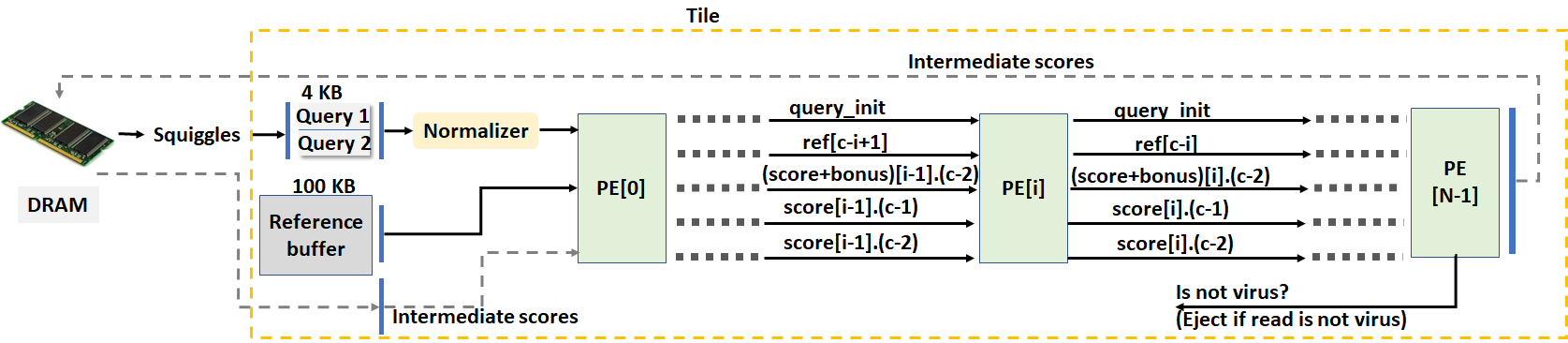}
	\vspace{-4mm}
  	\caption{\hwfilter~Tile. N=2000 PEs are connected with streaming inputs and outputs. The last PE determines the classification by comparing its cost to a threshold every cycle. $c$ is the cycle and $i$ is the PE index.}
  	\label{fig:array}
  	\vspace{-2mm}
\end{figure*}

\begin{figure}[h]
    \centering
	\includegraphics[width=8.5cm]{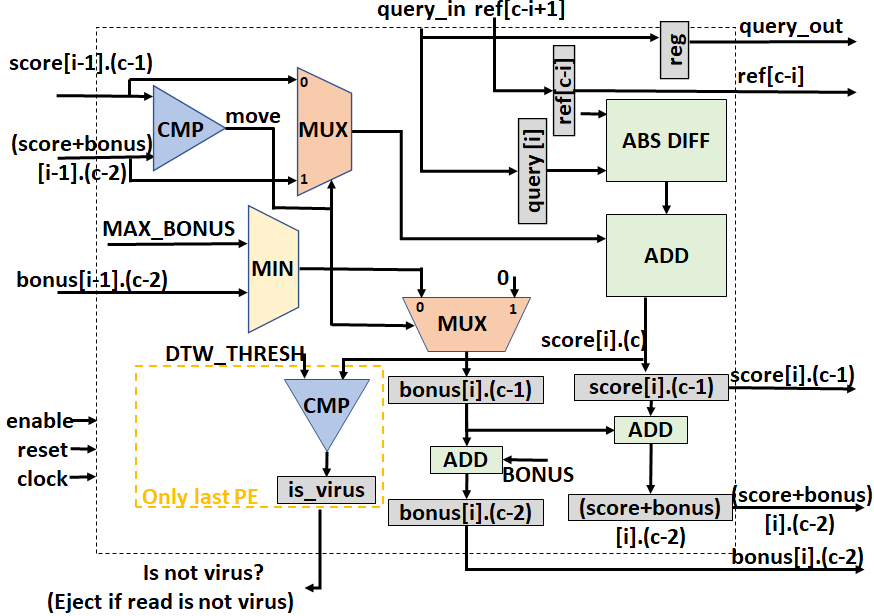}
	\vspace{-6mm}
  	\caption{\hwfilter~Processing Element.}
  	\label{fig:pe}
\end{figure}

\begin{figure}[h]
    \centering
	\includegraphics[width=8.5cm]{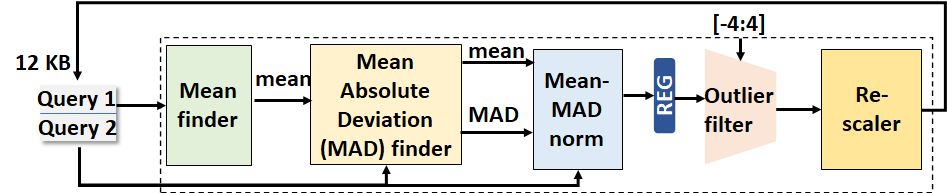}
    \vspace{-6mm}
  	\caption{\hwfilter~Normalizer. }
  	\label{fig:normalizer}
\end{figure}

\subsection{\hwfilter~Design}

\hwfilter~consists of 5 independent tiles (one tile is shown in Figure~\ref{fig:array}). Each can be individually power-gated based on desired filtering throughput. This number was chosen to meet the expected $100\times$ future increase in sequencing throughput.
Each read is assigned to an available tile for classification. As a read is sequenced, squiggles from a MinION R9.4.1 flow cell are streamed into DRAM in real-time. From there, squiggles are fetched into a tile's query buffers. Two ping-pong query buffers enable simultaneous squiggle loading and normalization. 
Once the desired length of read prefix has been sequenced, the raw squiggles of a query are normalized and then stored across the processing elements connected in a 1D systolic array. 

Each tile also stores a copy of the precomputed reference signal (loaded from flash during an initialization phase) in a reference buffer. The reference samples are then streamed into the systolic array. The entire sDTW matrix is computed in a wavefront parallel manner as described in Section~\ref{sec:alg}. The final PE determines the final minimum alignment cost, and sends a control signal to the MinION to eject the read if the final cost exceeds a predetermined threshold. Non-ejected reads are sequenced in full and stored in memory.

The number of cycles required to classify a new read is the read prefix length (2000 samples) plus the reference genome length (60,000 samples for SARS-CoV-2). 

{\bf Reference Buffer:}  We chose to use a separate buffer (100 KB) for each tile, even though all the reference buffers across the tiles store the same information (viral genome's reference squiggles). This allows us to reduce access latency and provide sustained throughput to each tile with just one read port. The area cost of duplicating the references is negligible, as reference buffers constitute only 6.98\% of total tile area. 

Furthermore, our design is independent of reference length and limited only by the reference buffer size provisioned. By loading a new precomputed reference signal onto the on-board flash, \hwfilter~can easily be reprogrammed to detect a novel virus.

{\bf Variable Query Length:}  As discussed in Section~\ref{multi}, there exists a trade off between classification accuracy and sequencing length of queries. We find (Section \ref{sec:savings}) that read prefix length of 2000 samples yields the most savings using Read Until, when we use a single threshold. Therefore, we use a 1D systolic array of size 2000 PEs. 

Our \hwfilter~design can handle variable read prefix lengths that are multiples of 2000 squiggle samples. To support query lengths longer than 2000 samples and multi-stage filtering, we configure the last PE such that it can optionally write the sDTW costs every cycle to DRAM. This consumes significant memory bandwidth. However, it enables sDTW computation to continue if greater classification accuracy by analyzing a longer prefix is desired. These intermediate costs are then loaded from DRAM and used to initialize the PEs (similar to initial normalized query) prior to computing the costs for a 4000-sample prefix length.

\subsection{Processing Element}
Each PE computes a cell in the sDTW matrix every cycle, using the final algorithm described in Section \ref{sec:alg}. At cycle $c$, each PE (Figure~\ref{fig:pe}) checks for the minimum among its previous neighbor's $c-1$ and $c-2$ cycle's outputs, modified by a bonus which rewards matching new reference bases. This minimum is then added to the absolute difference of the current query and reference values. Each PE stores the resulting costs and bonuses from its last two cycles for the next PE. Additionally, the last PE contains logic to compare its cost to a predefined threshold which determines whether or not to eject the read. This threshold can be reprogrammed on the SquiggleFilter based on software analysis of the target strain, but we have found it to be relatively robust across species and sequencing runs. Each PE is 1203$\mu\text{m}^{\text{2}}$ and requires 1.92mW when synthesized for a 28nm TSMC chip.

\subsection{Normalizer}
Normalization rescales the raw signals in order to improve classification accuracy when performing sDTW~\cite{sdtwax}, as discussed in Section~\ref{sec:norm}.
The normalizer, shown in Figure~\ref{fig:normalizer}, is a query preprocessor which streams in 10-bit samples from the query buffer for accumulation. After every $n=2000$ samples, the normalizer updates the mean and Mean Absolute Deviation (MAD), defined as follows: 
\begin{gather*}
    \mathrm{mean} = \bar{x} = \sum_{i=1}^n \frac{x_i}{n}\hspace{1cm} MAD = \sum_{i=1}^n \frac{|x_i-\bar{x}|}{n}
\end{gather*}
Thereafter, the streamed-in samples are transformed with mean-MAD normalization. The output normalized value is filtered for outliers and then re-scaled to a reduced precision 8-bit integer which is then fed to the tiles for sDTW classification. We find that 8 bits of precision is sufficient for accurate classification (Figure \ref{fig:alg_mods}). For efficiency, we do not convert the ADC sample to floating point, but instead use fixed-point values in the range $[-4,4]$.

\section{Methodology}
\label{methodology}

Human DNA datasets containing MinION R9.4 and R9.4.1 flow cells were obtained from the Nanopore Whole-Genome Sequencing Consortium~\cite{wgs_consortium} and the ONT Open Datasets~\cite{ont_open_data}. The SARS-CoV-2 dataset contains raw MinION R9.4.1 data available from the Cadde Centre~\cite{cadde}. We sequenced lambda phage DNA in our own laboratory using the ONT Rapid Library Preparation Kit~\cite{rapid_library_prep} following the Lambda Control protocol with a MinION R9.4.1 flow cell.

\sloppypar{We performed basecaller profiling measurements using a Titan XP GPU (server class) and Jetson Xavier GPU (edge class). Their specifications are provided in Table~\ref{tab:devices}. We evaluated both Guppy (\texttt{dna\_r9.4.1\_450bps\_hac.cfg}) and Guppy-lite (\texttt{dna\_r9.4.1\_450bps\_fast.cfg}) without modification using Guppy version 4.2.2~\cite{wick2019}. MiniMap2 version 2.17-r954-dirty~\cite{li2018minimap2} aligned basecalled reads.}

First, we measured the basecalling throughput of Guppy and Guppy-lite on a dataset of 33,004 full-length reads. Next, we used the proprietary Python libraries \texttt{ont-fast5-api} version 3.1.6~\cite{ont_fast5_api} and \texttt{ont-pyguppy-client-lib} 4.2.2~\cite{ont_pyguppy} to basecall the same reads in chunks of 2000 signals, thereby simulating Read Until on the same dataset. The Python code was instrumented to record latency information, and we tuned the number of reads simultaneously in-flight to optimize performance. This online Read Until processing (due to smaller batch size) resulted in 4.05$\times$ lower throughput for Guppy-lite and 2.85$\times$ lower throughput for Guppy on the Titan XP. Using these measurements and the relative peak throughputs of the Jetson and Titan, the Read Until performance of the Jetson Xavier was estimated (necessitated by the unavailability of \texttt{ont-pyguppy-client} ARM binaries for fine-grained Read Until control on the Jetson).

\begin{table}[htb]
  \centering
  \small
  \renewcommand*{\arraystretch}{1.2}
  \begin{tabular}{rcccc}
  \toprule
  & \textbf{Edge GPU} & \textbf{Edge CPU} & \textbf{GPU} & \textbf{CPU} \\
    \midrule
    \textbf{Model} & Jetson AGX & ARMv8.2 & Titan XP & 2$\times$ Intel Xeon \vspace{-2mm}\\
     & Xavier & & & E5-2697v3 \\
   \textbf{Cores} & 512 Volta & 8 & 3840 Pascal & 56 \\
   \textbf{Clock} & 1377MHz & 2265MHz & 1582MHz & 2600MHz \\
   \bottomrule
   \end{tabular}
  \caption{Architectural specifications of evaluated GPUs.}
  \label{tab:devices}\vspace{-2mm}
\end{table}

A memory-efficient multi-threaded implementation of sDTW was written in Python for accuracy analysis, and tested on 1000 reads from each of the datasets mentioned above. In order to determine the relative benefits of Read Until using different classification latencies and accuracies, we developed an analytical model to estimate sequencing runtime. This model accounts for factors such as average read length, desired coverage of the reference genome, average DNA capture time, and the Read Until parameters mentioned previously.

The design was first functionally verified via emulation on Amazon Web Service's EC2 F1 instance, which uses a 16nm Xilinx UltraScale+ VU9P FPGA. Further, \hwfilter~was synthesized using the Synopsys Design compiler for 28nm TSMC HPC and the design is clocked at 2.5GHz. 32GB 256-Bit LPDDR4x is connected to the System-on-Chip along with an 8-core ARM v8.2 64-bit CPU.

\ignore{
\begin{figure}[h]
    \centering
	\includegraphics[width=5cm]{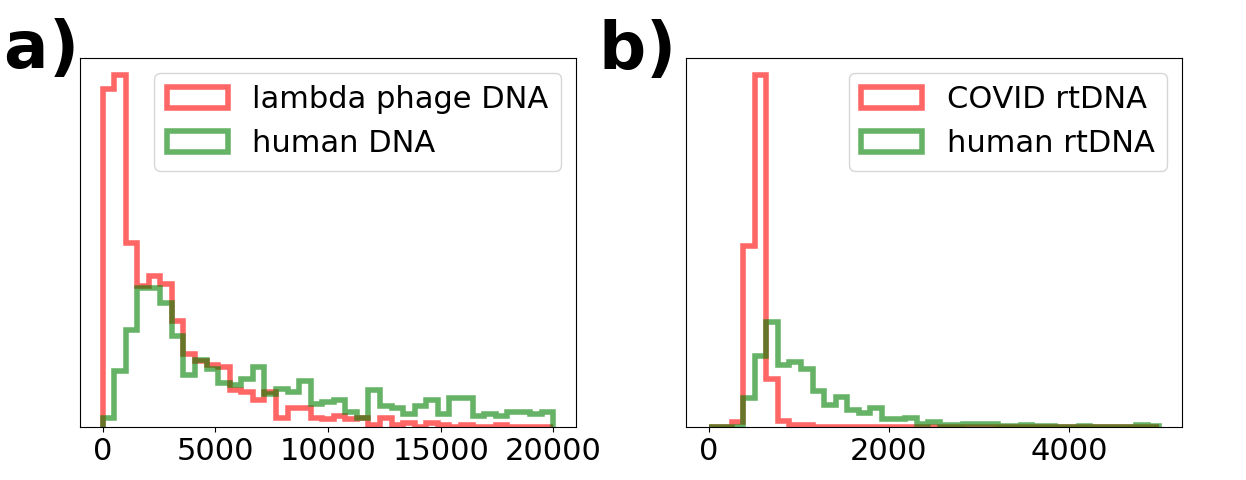}
  	\caption{Read length histograms (in base pairs) for the \textbf{a)} lambda phage-human DNA data and \textbf{b)} COVID-human rtDNA data.}
  	\label{fig:readlens}
\end{figure}
}

\section{Results}
\label{results}

\subsection{\hwfilter~Hardware Synthesis}
\label{synthesis_results}

\begin{table}[htb]
  \centering
  \begin{tabular}{ccc}
    \toprule
    \textbf{ASIC Element}  & \textbf{Area ($\text{mm}^{\text{2}}$)}  & \textbf{Power (W)}  \\
   \midrule
    Normalizer & 0.014 & 0.045 \\
    Processing Element & 0.001 & 0.002 \\
    Tile (1$\times$2000 PEs) & 2.423 & 2.780 \\
    Query buffer & 0.023 & 0.009 \\
    Reference buffer & 0.185 & 0.028 \\
    \midrule
    Complete 1-Tile ASIC & 2.65 & 2.86 \\
    Complete 5-Tile ASIC & 13.25 & 14.31 \\
   \bottomrule
   \end{tabular}
  \caption{\hwfilter~ASIC synthesis results.}
  \label{tab:synthesis_results}\vspace{-2mm}
\end{table}

\ignore{
\begin{table}[htb]
  \centering
  \scriptsize
  \begin{tabular}{cccc}
    \toprule
    \textbf{}  & \textbf{Max sequencer}  & \textbf{SquiggleFilter} & \textbf{Power (W)}  \\
    \textbf{}  & \textbf{throughput}  & \textbf{throughput} & \textbf{} \\
     \textbf{}  & \textbf{($\text{M samples/s}$)}  & \textbf{(M samples/s)} & \textbf{}\\
   \midrule
  \multicolumn{5}{l}{\textbf{Current}}   \\
  MinION &  2.05 & 80.64 & 3.19 \\
  GridION &  10.24 & 80.64 & 3.19 \\
  PromethION-24 &  288 & 415.28 & 14.31 \\
  \midrule
 \multicolumn{5}{l}{\textbf{Future}}   \\
  MinION &  204.8 & 403.2 & 14.31 \\
   \bottomrule
   \end{tabular}
  \caption{SquiggleFilter can be configured to handle current and future sequencer throughput. Numbers are for a 60Kbp reference and 2000-sample queries.}
  \label{tab:synthesis_results}
\end{table}
}

Table \ref{tab:synthesis_results} shows \hwfilter~synthesized to a 13.25$\text{mm}^{\text{2}}$ ASIC that consumes 14.21W when performing single-stage filtering and clocks at 2.5GHz. It contains 5 fully-independent tiles (which could be individually power-gated to improve energy efficiency). The latency for classifying a 2000-sample read from SARS-CoV-2 is 0.027ms, and for lambda phage is 0.043ms, due to its longer reference genome. This adds insignificant latency to each Read Until decision's critical path, since it takes around 500ms to sequence a sufficient number of bases to make an accurate decision. The single-tile classification throughputs for SARS-CoV-2 and lambda phage are 74.63M samples/s and 46.73M samples/s respectively, which are both considerably higher than MinION's current maximum output of 2.05M samples/sec). 
Additionally, if each tile is configured to perform multi-stage filtering, it will write intermediate results to DRAM, consuming only 10 GB/s main memory bandwidth per tile. Since Jetson Xavier's main memory supports 137 GB/s, our 5 tile design is feasible.

\subsection{Performance Analysis}
\label{sec:perf}

\begin{figure*}[h!]
    \centering
	\includegraphics[width=17cm]{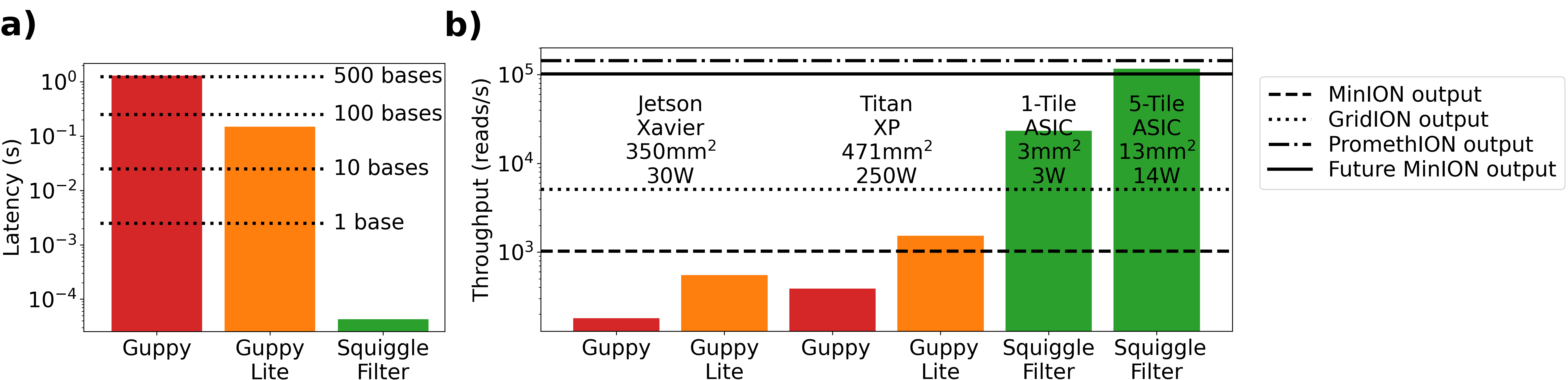}
	\vspace{-2mm}
  	\caption{\textbf{a)} Latency, and \textbf{b)} throughput of Guppy, Guppy-lite and \hwfilter~during Read Until.}
  	\label{fig:lat-thru}
  	\vspace{-2mm}
\end{figure*}

{\bf Latency:} Figure \ref{fig:lat-thru}a compares GPU-based basecalling latency to our \hwfilter~accelerator's latency. Note that we show only basecalling latency as it is the most time consuming step (96\% of compute time) of the virus classification pipeline. The measurements demonstrate that it would be impractical to use the high-accuracy Guppy basecaller as its latency is greater than one second, in which time more than 400 bases would have been unnecessarily sequenced for non-target reads. We found that Guppy-lite provides sufficient accuracy for Read Until classification as downstream aligner MiniMap2 is able to account for incorrect basecalls when aligning reads. However, a 149ms basecalling latency for Guppy-lite translates to an additional 60 bases sequenced for each read during classification. Since most non-target reads can be discarded after around 200 bases, this overhead is significant. In comparison, the common-case 0.04ms decision latency of \hwfilter~ensures that not even a single base pair is unnecessarily sequenced. 


{\bf Throughput:} Figure \ref{fig:lat-thru}b compares the basecalling throughput of Guppy-lite measured over GPU configurations to \hwfilter~accelerator's classification throughput.
An edge GPU such as the Jetson does not have sufficient compute power to basecall data from all pores in real-time and keep up with the maximum sequencing throughput of the MinION. We calculated that the Jetson's throughput would be approximately 95,700 bases per second, which is only 41.5\% of the MinION's maximum output of 230,400 bases per second. In the worst case, Read Until can only be performed using 41.5\% of the MinION's pores when basecalling using Guppy-lite on the Jetson. The remaining 59.5\% of pores are unable to use Read Until, and will sequence full-length human reads. In contrast, \hwfilter's throughput far exceeds MinION's and GridION's sequencing throughputs.

\subsection{sDTW Algorithm Accuracy}
\label{sec:acc}

\begin{figure*}[h]
    \centering
	\includegraphics[width=18cm]{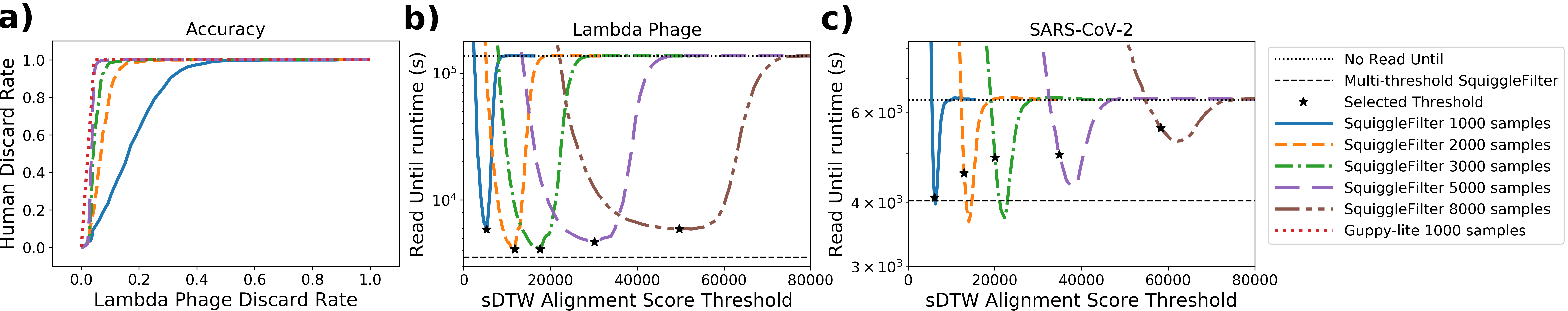}
	\vspace{-6mm}
  	\caption{SquiggleFilter Read Until \textbf{a)} accuracy, and performance on \textbf{b)} lambda phage and \textbf{c)} SARS-CoV-2 datasets.}
  	\label{fig:read_until}
  	\vspace{-3mm}
\end{figure*}

Figure \ref{fig:read_until}a compares sDTW accuracy to basecalling and alignment on a dataset of 1000 lambda phage and 1000 human reads, with a line plotted for each prefix length. The MiniMap2 alignment quality and sDTW alignment cost thresholds (for determining which reads to sequence and which to reverse) are swept through the range of possible values to show threshold-dependent accuracies. Although the Read Until accuracy obtained by basecalling and aligning slightly outperforms sDTW, this is to be expected since alignment algorithms such as MiniMap2 use numerous scoring heuristics and have matured significantly over the past two decades \cite{li2018minimap2}.

\begin{figure}[t]
    \centering
	\includegraphics[width=7.5cm]{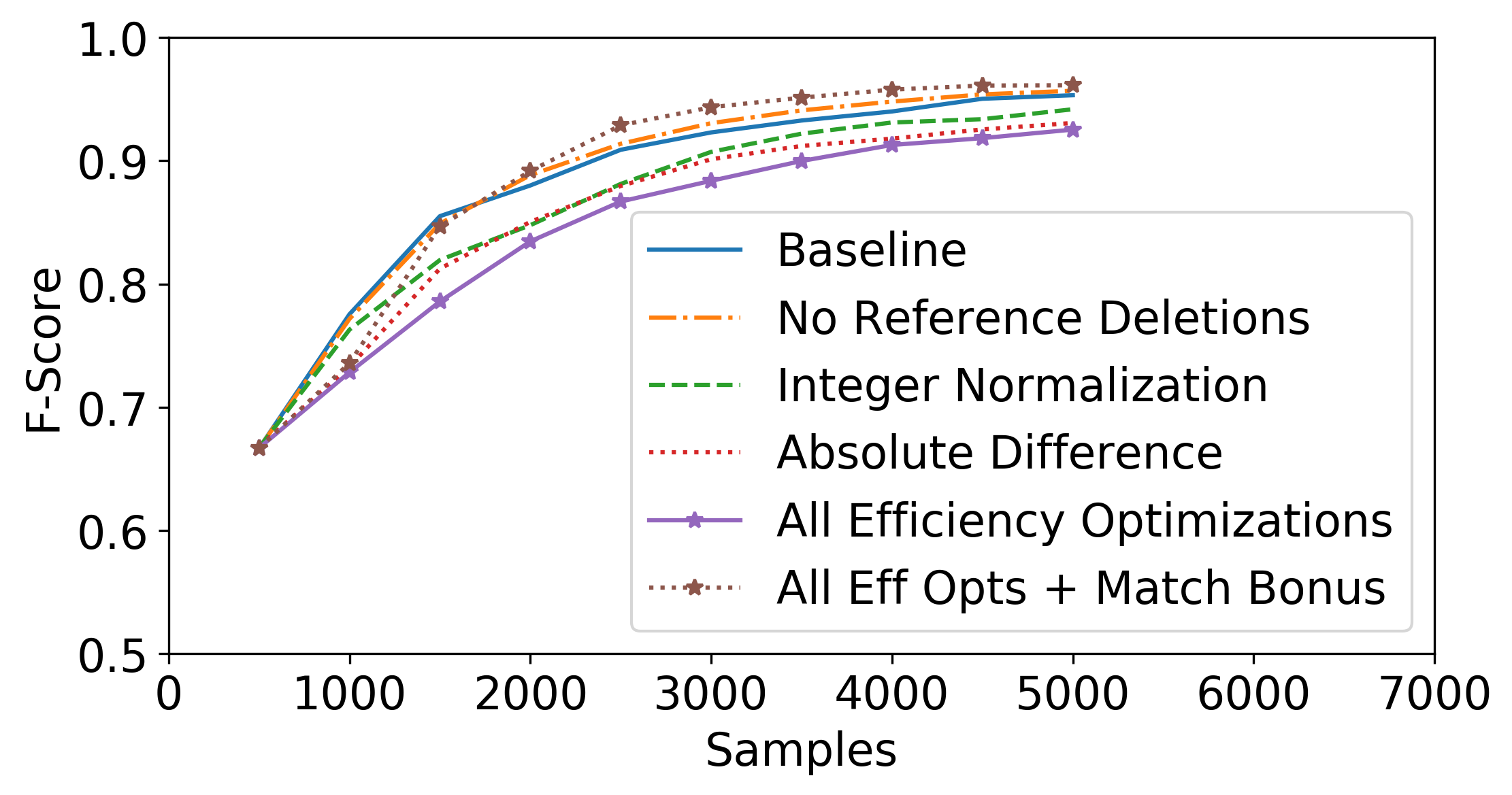}
	\vspace{-4mm}
  	\caption{Accuracy results for modifications to the standard sDTW algorithm.}
  	\label{fig:alg_mods}
  	\vspace{-2mm}
\end{figure}
\begin{figure}[h]
    \centering
	\includegraphics[width=7cm]{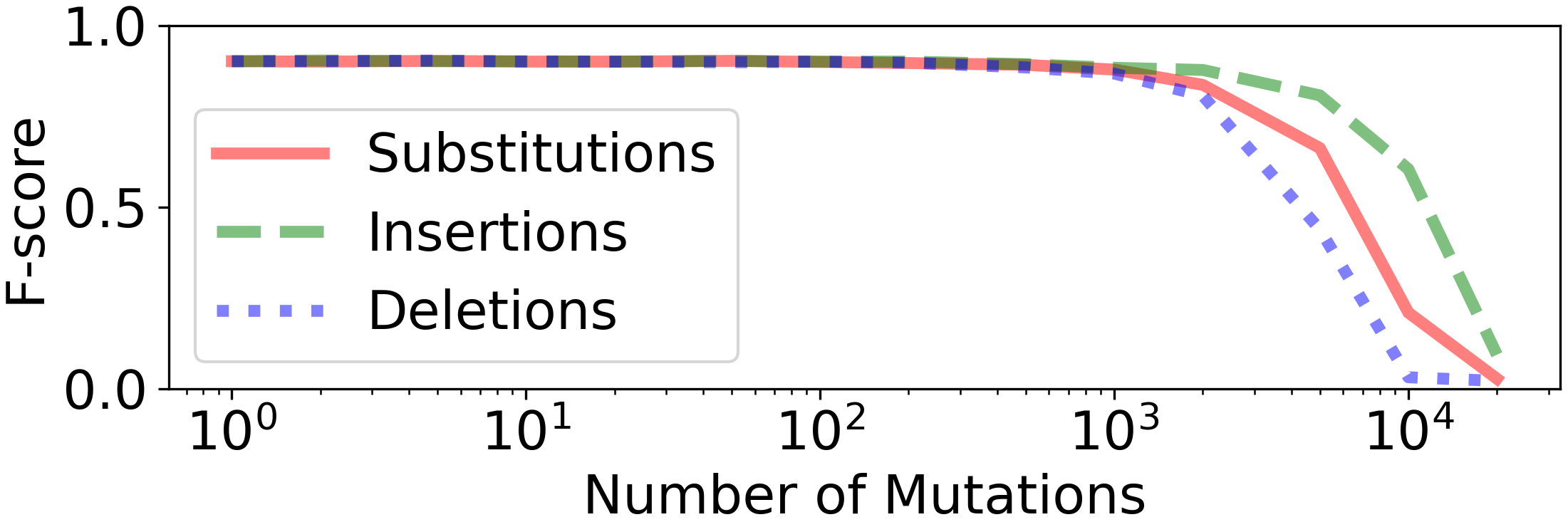}
	\vspace{-2mm}
  	\caption{SquiggleFilter accuracy is robust against random (lambda phage) reference mutations.}
  	\label{fig:mut}
  	\vspace{-2mm}
\end{figure}

Figure~\ref{fig:alg_mods} shows the maximal F-score for all of our algorithm modifications and standard sDTW on the same dataset. As expected, accuracy generally increases along with sample prefix length. 
We found that using both integer normalization and absolute difference for our distance metric reduce filtering accuracy slightly, a compromise which was expected. Eliminating reference deletions results in a slight accuracy improvement. 
Combining all three of these optimizations results in the lowest accuracy (but most efficient) of all configurations tested. We find that by including our ``match bonus'', we can recover lost accuracy and outperform the baseline, with a minor performance penalty. Figure~\ref{fig:mut} furthermore demonstrates that there is no a significant loss in filter accuracy until there is more than a 1,000 base difference between the reference genome and viral strain sequenced.

\subsection{Benefits of Read Until}
\label{sec:savings}

\begin{figure}[h]
    \centering
	\includegraphics[scale=0.40]{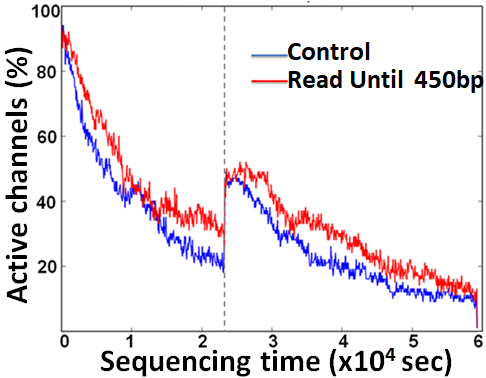}
	\vspace{-2mm}
  	\caption{Time saved is cost saved for sequencing. }
  	\label{fig:ru_time_cost}
  	\vspace{-3mm}
\end{figure}

Read Until not only saves sequencing time, but also cost. Figure~\ref{fig:ru_time_cost} shows our wet-lab experiment. After sequencing for a while, washing the flow cell with nuclease and re-multiplexing (rapid alternations of pore voltage bias direction, shown with dotted black line) leads to  control and Read Until pores having the same number of active channels. This means that Read Until does not damage the flow cell any more than normal sequencing, but enables more experiments to be run over the lifetime of any flow cell.


The single-threshold Read Until design space was first explored for our lambda phage dataset. Figure~\ref{fig:read_until}a shows the accuracy of \hwfilter~for a variety of Read Until prefix lengths (each line), and for all reasonable sDTW alignment cost thresholds (points on each line). Given this experimentally measured accuracy, the total expected sequencing time to perform Read Until for lambda phage was calculated using our analytical model, and is shown in Figure~\ref{fig:read_until}b. 
We found that the best single-threshold configuration for \hwfilter~outperforms Guppy-lite on this dataset by 12.9\% in terms of Read Until runtime. By using multiple thresholds, we can reduce runtime by a further 13.3\%. 

A similar analysis was then performed for the SARS-CoV-2 dataset, and the results are shown in Figure \ref{fig:read_until}c. Optimal sDTW alignment cost thresholds were taken from the Read Until runtime minima from Figure \ref{fig:read_until}b, and the corresponding Read Until runtimes using those thresholds are marked for the SARS-CoV-2 dataset. 


\subsection{Looking Forwards: Scalability}

\begin{figure}[h]
    \centering
	\includegraphics[width=8cm]{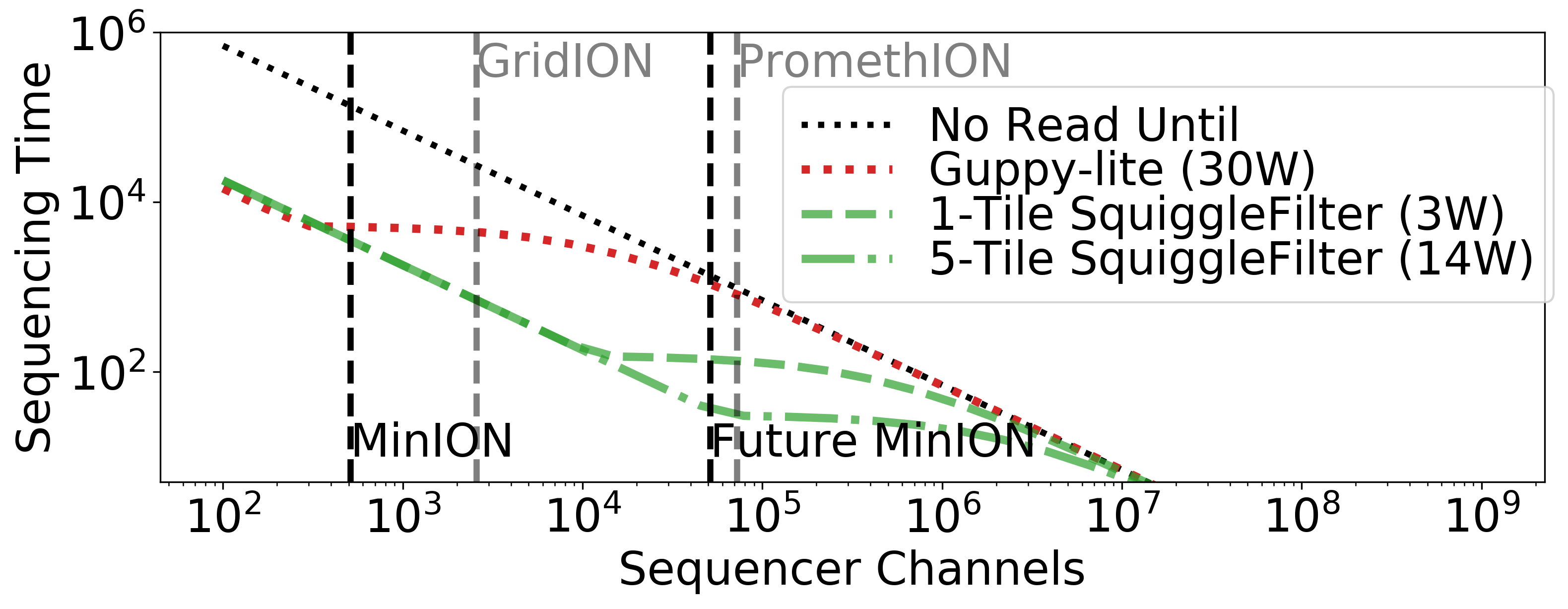}
	\vspace{-3mm}
  	\caption{Future \hwfilter~Read Until benefits.}
  	\label{fig:scale}
  	\vspace{-3mm}
\end{figure}

Sequencing throughput is expected to increase by $10-100\times$ within the next few years, due to new nanopore chemistry enabling a denser configuration with many more channels per flow cell~\cite{ont_tech_update}. Figure~\ref{fig:scale} shows that without further improvements to basecalling throughput, current GPUs will be unable to keep pace with new sequencing technology. As a result, the time and cost savings gained through Read Until will be largely lost. We can see that Guppy-lite's slight edge over \hwfilter~in terms of accuracy has already been lost due to its inability to perform Read Until on 512 pores. In contrast, our \hwfilter~accelerator can tolerate a $114\times$ increase in sequencing throughput.
\section{Related Work}
\label{related}


The MinION was released in 2014 as the first commercially available nanopore-based DNA/RNA sequencing device~\cite{minion}. The first Read Until software pipeline was developed two years later, in 2016~\cite{read_until_dtw}. In this seminal work, raw nanopore signal was first segmented into events, and then events were aligned to a lambda phage reference using subsequence Dynamic Time Warping (described in Section~\ref{sec:sdtw}). Event segmentation is used to detect the most likely positions in the raw signal where a new base has entered the pore, and could be considered a rudimentary form of basecalling. In fact, it has been used as an essential preprocessing step in several older basecallers~\cite{wick2019}. Unfortunately, the throughput measured by this original work on an 8-core ARM processor is 40$\times$ lower than the current maximum MinION output. 

As basecalling throughput and accuracy has gradually increased over the last few years, the standard approach for Read Until pipelines has been to basecall the signal and use an aligner to determine if each read aligns to the target genome~\cite{rubric_new, rubric, read_until_api, ru_lec}. This method achieves the highest accuracy, but is not scalable. When pairing a server-class GPU with a handheld MinION device, it is just able to perform Read Until with the required throughput, albeit with significant latency (as shown in Section~\ref{sec:perf}).

UNCALLED, a more recent work, skips basecalling by doing approximate alignments in 3 steps: event segmentation, FM-index look-ups, and seed clustering~\cite{kovaka}. However, we evaluated UNCALLED and observed that it requires longer prefix lengths for accurate alignment. 23.63$\%$ of 2000-sample long chunks from our lambda phage dataset were not alignable. After segmentation, UNCALLED uses an FM-index to filter reads. UNCALLED aligns only $\sim$76$\%$ of the lambda reads of 2000 samples on a modern Intel i7-7700 desktop processor taking 16ms per read. Moreover, $\sim$14$\%$ of reads take ~353ms per read to be aligned as more samples are required for a decision. $\sim$10$\%$ of the reads, however, are left unaligned. On an edge device with an ARM core and lower memory bandwidth, performance would be worse. No existing software-only solution has adequate throughput and low enough latency to effectively perform Read Until on an edge device.

In contrast, our approach shifts to a minimalistic sDTW alignment algorithm, and by designing hardware to accelerate the simple and regular sDTW computation, we can easily meet the desired throughput and latency requirements on an edge device. General purpose DTW accelerators have already been designed to solve alignment problems in other domains such as audio signal processing~\cite{vlsi_dtw_ax} and astronomy~\cite{sdtwax}, but nanopore viral DNA/RNA filtering required several application-specific optimizations to meet the desired latency, throughput and accuracy requirements. Our design involves several algorithmic modifications to vanilla sDTW (described in Section~\ref{sec:alg}), uses an on-chip buffer for efficient repeated alignments to the same reference, replaces all floating-point computation with integer arithmetic for increased efficiency, uses multi-stage filtering for optimal Read Until results, and has been evaluated on a novel virus (SARS-CoV-2).

There has recently been significant work on designing hard-ware accelerators for genomics applications~\cite{genax, genesis, darwin,seedex,gencache,nvm_read,genasm,wu2019fpga}, but these accelerators focus on human genome sequencing. As a result, they efficiently align many (usually short) basecalled reads to a long reference genome with high throughput and accuracy. As noted previously in Section~\ref{sec:need}, our problem has very different computational needs. We must selectively filter short noisy raw signals (squiggles) with sufficiently high throughput and low latency to effectively exploit Read Until. We achieve this by replacing the basecaller and aligner with \hwfilter. 

\section{Conclusion}	
\label{conclusion}

In designing a universal virus detector, we identify the basecaller to be a significant bottleneck in filtering non-target reads. This compute problem is only going to get worse, as the throughput of nanopore sequencers is expected to increase by 10-100$\times$ in the near future. We address this problem using hardware-accelerated \hwfilter~for filtering non-target reads without basecalling them.
We show that our 14.3W 13.25$\text{mm}^{\text{2}}$ accelerator has 274$\times$ greater throughput and 3481$\times$ lower latency than existing approaches while consuming half the power, enabling Read Until for the next generation of nanopore sequencers.
\begin{acks}
We thank Robert Dickson and John Erb-Downward for introducing us to targeted nanopore sequencing for clinical diagnostics and for borrowed use of their Jetson AGX Xavier. We also thank Jenna Wiens, Piyush Ranjan, Arun Subramaniyan, and Yichen Gu for their helpful input and feedback at various stages of this project. Lastly, we would like to thank the ONT community as a whole.
\end{acks}
%
%
%
%
%
\clearpage
\appendix
\section{Artifact Appendix}

\subsection{Abstract}
Our artifact contains the RTL and testbench SystemVerilog code for our SquiggleFilter accelerator in the \texttt{design/} subdirectory. Additionally, \texttt{sdtw\_analysis.ipynb} is a full Jupyter Notebook pipeline containing our software sDTW algorithm implementation and our Read Until runtime model, along with scripts for generating multiple figures from our paper.

\subsection{Artifact check-list (meta-information)}

\begin{itemize}\setlength\itemsep{0mm}
  \item {\bf Algorithm:} Hardware and software implementation of custom subsequence Dynamic Time Warping (sDTW) algorithm for filtering non-viral DNA reads in real time.
  \item {\bf Program:}  RTL and SystemVerilog testbench code for \hwfilter~accelerator. Jupyter Notebook containing Python sDTW implementation and runtime model.
  \item {\bf Data set:} Raw human, lambda phage, and SARS-CoV-2 FAST5 data from several public sources~\cite{cadde,ont_open_data}.
  \item {\bf Run-time environment:} Vivado 2019.1 and Jupyter Notebook. Build instructions targeted to Ubuntu 18.
  \item {\bf Hardware:} At least one CPU core and 10GB RAM for the notebook. Recommended requirements for Xilinx Vivado based on Xilinx SDK: min 2.2GHz, Intel Pentium 4, Intel Core Duo, or Xeon Processors; SSE2 minimum.
  \item {\bf Output:} Software regeneration of multiple figures from the paper. Verification of hardware using SystemVerilog testbench.
  \item {\bf How much disk space required (approximately)?:} 40GB public dataset download. 40GB for public dataset download. Xilinx Vivado requires upto 30GB of diskspace for installation and an additional 2.5GB if Vivado simulation is started. 
  \item {\bf How much time is needed to complete experiments (approximately)?:} Jupyter Notebook requires 10 minutes with 56 cores. Vivado simulation on the SARS-CoV-2 reads can take 1-21 minutes on a Quadcore 8th Gen i5 with 8GB RAM depending on the number of test-cases anyone may wish to run.
  \item {\bf Publicly available?:} Yes.
  \item {\bf Archived (provide DOI)?:} \url{https://doi.org/10.5281/zenodo.5150973}
\end{itemize}

\subsection{Description}

\subsubsection{How to access}
All of the source code is open source, and can be obtained either through GitHub\footnote{\label{github}https://github.com/TimD1/SquiggleFilter} or Zenodo\footnote{\label{zenodo}https://doi.org/10.5281/zenodo.5150973}.

\subsubsection{Hardware dependencies}
The \hwfilter~code requires approximately 10GB of RAM, and the datasets used require approximately 40GB of disk space. Xilinx Vivado comes with the following additional requirements on the processor: minimum 2.2GHz, Intel Pentium 4, Intel Core Duo, or Xeon Processors; SSE2 minimum. 

\subsubsection{Software dependencies}
Any Linux OS can be used, but a recent Ubuntu release is recommended for ease of installation. The Jupyter Notebook has multiple Python package dependencies, which will be installed by the \texttt{setup.sh} script. For hardware evaluation, a recent installation of the licensed Vivado Design Suite is recommended; we used release 2019.1. Further details on the installation can be found on \url{ https://www.xilinx.com/support/download/index.html/content/xilinx/en/downloadNav/vivado-design-tools/archive.html}.

\subsubsection{Data sets}
Our artifact uses three raw nanopore signal (FAST5) datasets:
\begin{itemize}\setlength\itemsep{0mm}
    \item \textbf{lambda}: This dataset of 21,000 lambda phage reads was generated in our laboratory, and is included in our GitHub repository at \texttt{data/lambda/fast5}.
    \item \textbf{covid}: This dataset of 1.2 million SARS-CoV-2 reads is downloaded from the CADDE Centre~\cite{cadde} to \texttt{data/covid/fast5} by the \texttt{setup.sh} script.
    \item \textbf{human}: This dataset of 65,000 huan reads is downloaded from ONT Open Datasets~\cite{cadde} to \texttt{data/human/fast5} by the \texttt{setup.sh} script.
\end{itemize}

\subsection{Installation}

All source code is available in either our GitHub$^1$ or Zenodo$^2$ repositories.
\begin{itemize}\setlength\itemsep{0mm}
    \item \textbf{README.md} contains instructions for evaluating the artifacts
    \item \textbf{design/} contains the SystemVerilog RTL and testbench. \texttt{testbench\_top.sv} is the top file of the testbench for behavioral simulation. \texttt{normalizer\_top.v} is the top file for the normalizer and it's sub-modules. \texttt{warper\_top.sv} is the top file for the systolic array.
    \item \textbf{sdtw\_analysis.ipynb} contains our software pipeline, Python sDTW implementation, and runtime model. 
    \item \textbf{setup.sh} is the setup script
    \item \textbf{data/} contains all three datasets
    \item \textbf{scripts/} contains all scripts used for data analysis
\end{itemize}
Please follow all instructions from \texttt{README.md} to evaluate the artifacts.


\subsection{Evaluation and expected results}
\subsubsection{Hardware}
After installing and running Vivado, go under settings and change the simulation run time to 18ms for complete simulation. On the flow navigator, pressing the run simulation option would start the simulation and messages would start appearing on the tcl console printing whether the testcases passed or failed. We observe and expect all the testcases to pass. Additionally, the waveform may be viewed as the simulation begins. Please find detailed instructions in \texttt{README.md}.

\subsubsection{Software}
After the Jupyter Notebook is running, please select the \texttt{sf-venv3} kernel (\texttt{Kernel -> Change Kernel}) created by the \texttt{setup.sh} script. Then, run all cells in order (\texttt{Kernel -> Restart and Run All}). The entire pipeline should run successfully, computing the sDTW scores on the datasets selected and regenerating most of the figures in our paper.



\newpage
\subsection{Methodology}

Submission, reviewing and badging methodology:
\vspace{-2mm}
\begin{itemize}\setlength\itemsep{0mm}
  \item \url{https://www.acm.org/publications/policies/artifact-review-badging}
  \item \url{http://cTuning.org/ae/submission-20201122.html}
  \item \url{http://cTuning.org/ae/reviewing-20201122.html}
\end{itemize}

\bibliographystyle{ACM-Reference-Format}
\bibliography{references}
\end{document}